\documentclass[conference]{IEEEtran}

\usepackage{url}
\usepackage{array}
\usepackage{multirow}
\usepackage[labelformat=simple]{subcaption}
\usepackage[linesnumbered,longend,ruled,vlined]{algorithm2e}
\usepackage{fixltx2e}
\usepackage{amssymb}
\usepackage{amsmath}
\usepackage{cite}
\usepackage{float}
\usepackage{tabularx,colortbl}
\ifCLASSINFOpdf
  \usepackage[pdftex]{graphicx}
\else
  \usepackage[dvips]{graphicx}
\fi
\usepackage{tikz}

\newcommand*\circled[1]{\tikz[baseline=(char.base)]{
            \node[shape=circle,draw,inner sep=1pt] (char) {#1};}}

\let\oldnl\nl
\newcommand{\nonl}{\renewcommand{\nl}{\let\nl\oldnl}}
\newcolumntype{?}{!{\vrule width 0.5pt}}
\newcommand*{\qed}{\hfill\ensuremath{\square}}

\newtheorem{definition}{Definition}
\newtheorem{example}{Example}

\newtheorem{property}{Property}

\begin{document}


\title{An Efficient Summary Graph Driven Method for RDF Query Processing}

\author{
\IEEEauthorblockN{Lei Gai,
Wei Chen and
Tengjiao Wang
}
\IEEEauthorblockA{\IEEEauthorrefmark{1}School of Electrical Engineering and Computer Science\\
Peking University. Beijing, P.R.China\\
Email: \{lei.gai, pekingchenwei, tjwang\}@pku.edu.cn}
}

\maketitle

\begin{abstract}
RDF query optimization is a challenging problem. Although considerable factors and their impacts on query efficiency have been investigated, this problem still needs further investigation. We identify that decomposing query into a series of light-weight operations is also effective in boosting query processing. Considering the linked nature of RDF data, the correlations among operations should be carefully handled. In this paper, we present SGDQ, a novel framework that features a partition-based \textit{Summary Graph Driven Query} for efficient query processing. Basically, SGDQ partitions data and models partitions as a summary graph. A query is decomposed into subqueries that can be answered without inter-partition processing. The final results are derived by perform summary graph matching and join the results generated by all matched subqueries. In essence, SGDQ combines the merits of graph match processing and relational join-based query implementation. It intentionally avoids maintain huge intermediate results by organizing sub-query processing in a summary graph driven fashion. Our extensive evaluations show that SGDQ is an effective framework for efficient RDF query processing. Its query performance consistently outperforms the representative state-of-the-art systems.
\end{abstract}

%
\IEEEpeerreviewmaketitle

%
%
%
%

\section{Introduction}\label{sec-introduction}

The Resource Description Framework(RDF) have been pervasively adopted in many fields, such as knowledge bases(DBpedia\cite{DBpedia}, YAGO\cite{YAGO}, Freebase\cite{freebase}), semantic social networks (FOAF\cite{foaf}, Open Graph\cite{opengraph}) and bioinformatics(Bio2RDF\cite{bio2rdf}, UniProt\cite{uniprot}), etc. The available RDF data keep increasing in both size and quality, this challenges for a framework for efficient query processing. For schema-flexible RDF data, query is naturally a graph pattern matching process. Generating results in a refined data space can greatly boost the RDF query performance. To this end, a considerable exploratory methods have been proposed in the recent decade.

One class of solutions utilize the graph nature of RDF data and implement query as subgraph matching\cite{gstore-vldb2011,trinitrdf-vldb2013,subgraph-vldb2015}. Subgraph matching only needs to explore the neighborhood data, and generates results in an incremental fashion. This avoids maintaining large intermediate results\cite{trinitrdf-vldb2013}. Nonetheless, the scale of RDF graph have a profound impact on query performance. The time complexity of subgraph matching increase exponentially as the graph size increase. Furthermore, as graph exploration is based on random data access. the efficiency of exploration-intensive subgraph match is guaranteed by holding the whole graph topology in memory, which is infeasible for large-scale graph. Recent researches\cite{structuralindex-tkde2013, triad-sigmod2014} introduced graph summarization in RDF query, The condense summary graph are solely used in filtering out irrelevant data in query.

The other class, which are the mainstream solutions, are based on 40+ years of relational database researches\cite{rdf3x2010, triplebit-vldb2013}. In principle, RDF data is managed as relational tables, and joins are excessively utilized in query implementation. The scan and join on large table are prone to introduce large intermediate results, which can easily  overwhelm the system memory and cause memory thrashing that greatly degrade query performance. Existing query optimization methods mainly focus on triple indexing techniques(e.g. \cite{permuindex-laweb2005,bitmat-www2010}) and query optimization methods(e.g. \cite{heuristics-www2008,complexquery-edbt2014}) that utilizing the typical characteristics of RDF. Using relational solutions, we argue that splitting RDF data into smaller parts that can be processed independently is also effective in optimizing query processing. On the flip side, there is no one-size-fit-all partition method, some results may span across partitions. The existence of inter-partition processing complicated the query implementation in this manner. Considering that the extensive connections among partitions can be naturally modeled using graph structure, this envisions us to introduce a summary graph and use subgraph match to drive the operations on partitions.

In this paper, we present a new RDF processing framework named SGDQ(\textit{\textbf{S}ummary \textbf{G}raph \textbf{D}irected \textbf{Q}uery}). SGDQ partitions data and decomposes a query into sub-queries, such that each sub-query can be answered independently in partitions. All sub-query processing are driven by a summary graph matching process, and the final results is generated as joins between subqueries in a match. The rational behind SGDQ is two folds. First, SGDQ alleviates the need of maintaining massive intermediate results by decomposing query process into a series of light-weight sub-queries. Second, subgraph match is effective in representing the inter-partition query processing, and it is can be efficiently implemented on a summarized graph of moderate scale.

To achieve this, two aspects in SGDQ need to be carefully considered. The first aspect is the design of physical data layout that support decomposed sub-query processing. We distinguish triples that do not contribute to graph partition and manage them in a specific manner. To facilitate summary match, we adopt a graph partition based strategy for data partition. Each partition is managed individually and the inter-connection among partitions is modeled as a summary graph. The second aspect is the efficient processing of sub-queries. We propose a process that use auxiliary patterns to filter out irrelevant data in sub-query processing, and adopt a mechanism to prevent duplications in final results.

We summarize our major contributions in this paper as follows:
\begin{itemize}
    \item We present SGDQ(\textit{\textbf{S}ummary \textbf{G}raph \textbf{D}irected \textbf{Q}uery}), a novel and effective framework for RDF query processing. SGDQ decomposes a query into a series of light-weight sub-queries, and leverages summary graph matching paradigm to drive the processing of sub-queries. SGDQ can greatly reduce the amount of data that need to maintain during query, and boost query performance by effective summary graph matching and efficient sub-query processing.
    \item We propose a specialized physical data layout that support SGDQ, and provide efficient operations based on this physical data layout. These operations are the building blocks of SGDQ.
    \item We perform extensive evaluations using two types of benchmarks. The evaluation results show that the query performance of SGDQ consistently outperforms three existing representative RDF systems.
\end{itemize}

The rest of the paper is organized as follows. \textit{Section} \ref{sec-preliminaries} describes the principle concepts.  \textit{Section} \ref{sec:overview} gives an overview of SGDQ framework. \textit{Section} \ref{sec-partition} presents the physical data layout and implementation of sub-query processing. \textit{Section} \ref{sec:sgdq} details the paradigm of summary graph driven processing. The most related works is reviewed in \textit{Section} \ref{sec-related} , and \textit{Section} \ref{sec-evaluation} presents the evaluation results. We conclude in \textit{Section} \ref{sec-conclusion}.

%
%
%
%
\begin{figure*}
\captionsetup{belowskip=0pt,aboveskip=0pt}
\captionsetup[sub]{margin=0.5pt,skip=0.5pt, labelfont=bf}
\begin{minipage}[b][0.34\textheight][s]{0.31\textwidth}
  \centering
  \includegraphics[height=0.30\textheight,width=\textwidth]{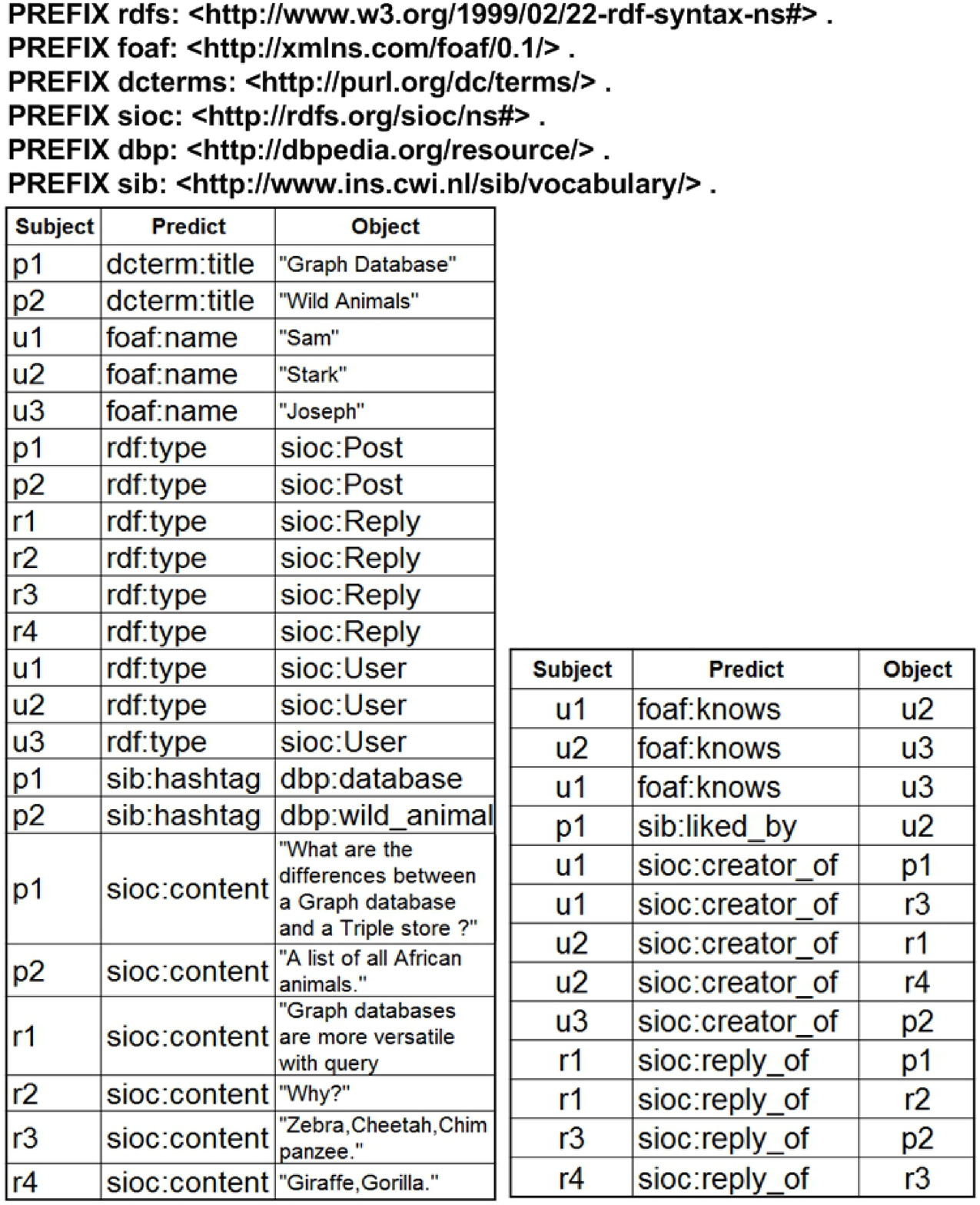}
  \subcaption{An examplar RDF dataset, $\mathcal{D}$.}
  \label{subfig:rdf-collection}
\end{minipage}%
\qquad
\begin{minipage}[b][0.34\textheight][s]{0.36\textwidth}
  \centering
  \includegraphics[height=0.30\textheight,width=\textwidth]{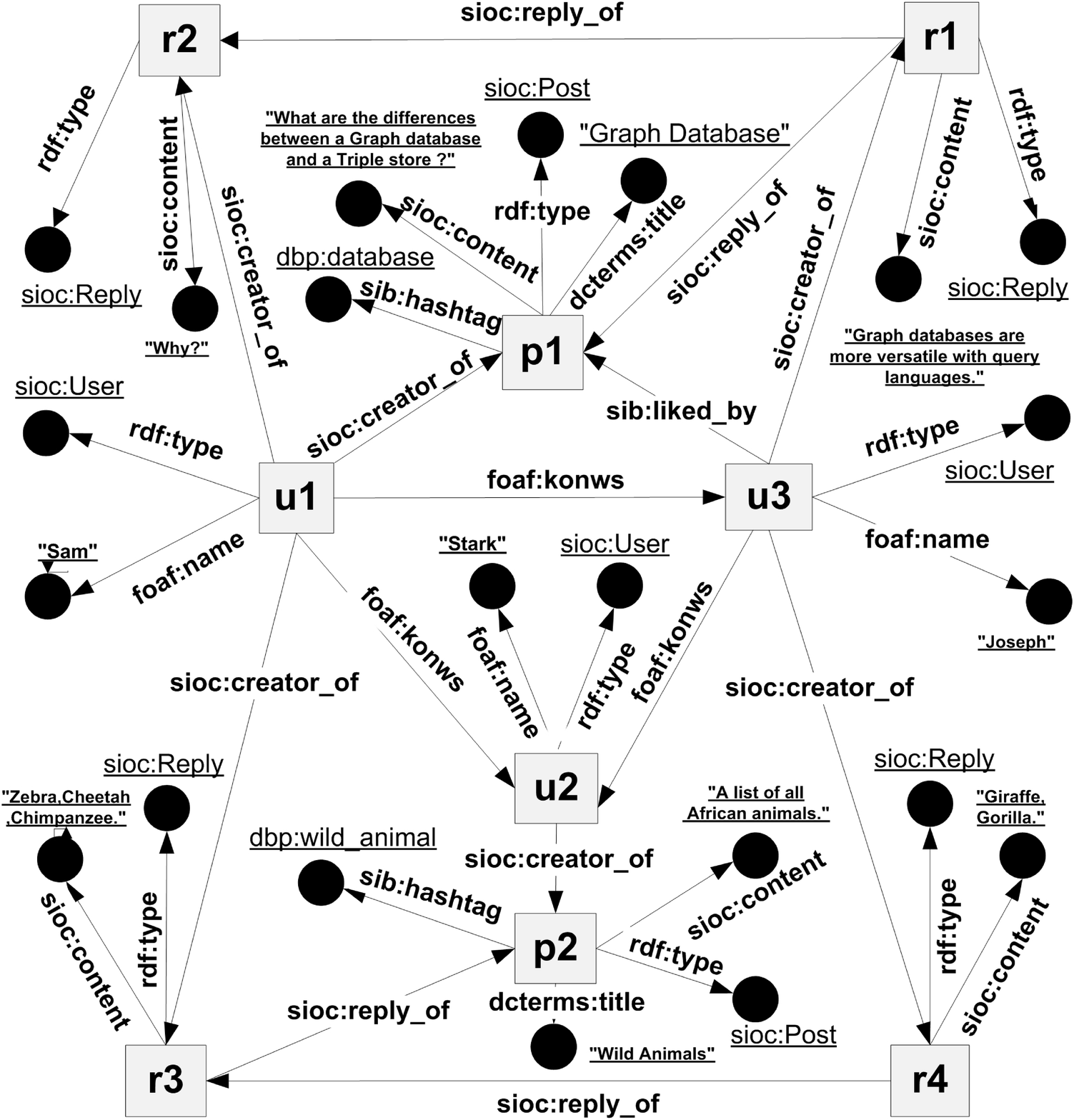}
  \subcaption{ $\mathcal{G}$, the RDF graph of $\mathcal{D}$.}
  \label{subfig:rdf-graph}
\end{minipage}%
\qquad
\begin{minipage}[b][0.34\textheight][s]{0.25\textwidth}
  \centering
  \vspace{\baselineskip}
  \includegraphics[height=0.25\textheight,width=1.0\textwidth]{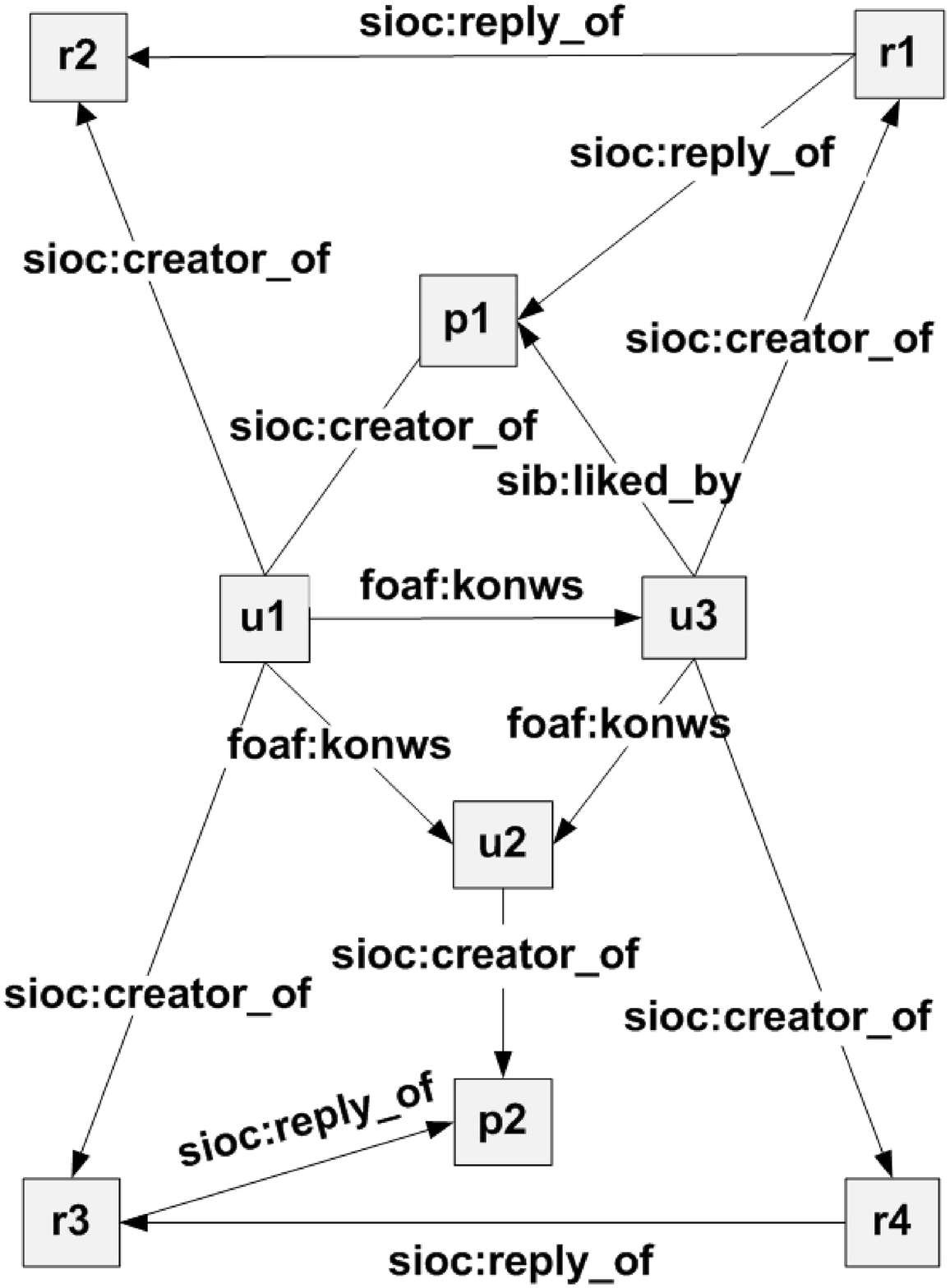}
  \vspace{\baselineskip}
  \subcaption{$\mathcal{G}_{R}$, the RL graph of $\mathcal{D}$.}
  \label{subfig:rl-graph}
\end{minipage}\hfill
\label{fig:data-example}
\caption{An exemplar RDF dataset $\mathcal{D}$, its data graph $\mathcal{G}$ and RL-graph $\mathcal{G}_R$ }
\end{figure*}

\section{Preliminaries}\label{sec-preliminaries}

RDF is a W3C recommended standard for representing and publishing linked web resources. Due to its simple and schema-free nature, RDF facilitates the representation and integration of data from different domains. In RDF, A real world \textit{resource} is identified by an Uniform Resource Identifier(URI) string. Its properties and linkages to other resources are described by a set of triples. Each triple is uniformly represented as three elements $\langle$\textit{Subject, Predicate, Object}$\rangle$, or $\langle$s,p,o$\rangle$ for brevity. We call \textit{p} a \textit{relation} if \textit{o} is a resource URI, or an \textit{attribute} if \textit{o} is a literal(string, number, or date, etc)\footnote{There are some exceptions. Following the semantics of RDF schema, predicate \textit{rdf:type} is widely used to link instances with classes. Thus in this paper, \textit{rdf:type} is classified as an \textit{attribute}. Furthermore, some predicates can also be \textit{attribute}. The selection of these \textit{attribute}-predicate is dataset specific. }. A collection of such triples forms an \textit{RDF dataset} $\mathcal{D}= \bigcup^{|\mathcal{D}|}_{i=1} t_i$. $\mathcal{D}$ can be naturally modeled as a directed, edge-labeled \textit{RDF graph} $\mathcal{G}=\langle V,E \rangle$. Each triple $t_i \in \mathcal{D}$ corresponds to a directed labeled edge $e \in E$, which connects vertex \textit{s} to vertex \textit{o} with an edge value \textit{p}.

Like SQL in relational database, users commonly resorts to declarative query for RDF data access. Adopting query language such as SPARQL\cite{SPARQL-url}, a conjunctive RDF query $Q$ can be expressed as: \textbf{SELECT} $?v_1,\ldots$ \textbf{FROM} $TP$, where $TP$ is a combination of triples, each one of which has a least one of \textit{s}, \textit{o} or \textit{p} replaced by variables $?v$. We refers to such triples as \textit{triple patterns}, and denotes a query as $Q=\{tp_1,\ldots, tp_k\}$. Two triple patterns are \textit{joinable} if they implicitly share a common elements. In this paper, we assume that all predicates of \textit{triple pattern}s may not be variables. This assumption is reasonable for that \textit{Predicate} join is infrequent in practice, as shown in a previous study\cite{realSparqlQuery}. This leads to 3 types of joins, named according to the position of common elements: \textit{s-s join}, \textit{o-o join}, \textit{s-o join}. By considering each $tp_i$ as an edge and follows same way of mapping $\mathcal{D}$ to $\mathcal{G}$, $Q$ can also be viewed as a \textit{query graph} $G^Q$.  Accordingly, a query can be implemented as the process of match $G^Q$ on $\mathcal{G}$.

The notations used in this paper are listed in \textit{Table} \ref{tbl:notations}. Next, we further introduce some principle concepts.
\begin{table}[ht]
\captionsetup{belowskip=0pt,aboveskip=0pt}
\caption{Notations used in this paper. }
\label{tbl:notations}
\scriptsize
\centering   
\begin{tabular}{ | >{\raggedright\arraybackslash}m{0.5in} | >{\raggedright\arraybackslash}m{2.6in}  |}
    \hline
    \multicolumn{1}{|c|}{\textbf{Notation}} &   \multicolumn{1}{c|}{\textbf{Description}}    \\
    \hline
    \hline
    $\mathcal{D}$                           &   RDF dataset, $\mathcal{D} = \bigcup_{i=1}^{|\mathcal{D}|} t_i$, $t_i$ in N-Triple (s,p,o).     \\
    \hline
    \textit{s}, \textit{p} and \textit{o}    &   denotes the value of \textit{Subject},\textit{Predicate} and \textit{Object} in a triple. \\
    \hline
    $\mathcal{D}_A, \mathcal{D}_R$           &   The set of \textit{A-Triples} and \textit{R-Triples}.     \\
    \hline
    $\mathcal{M}_A, \mathcal{M}_R$           &   The global dictionary of $\mathcal{D}_A$ and $\mathcal{D}_R$.     \\
\hline
    $\mathcal{G}$                           &   The RDF graph of $\mathcal{D}$, $\mathcal{G}=\langle V,E\rangle$.   \\
    \hline
    $Q$                                     &   Query, as a set of $k$ triple patterns. $Q=\{tp_1,\ldots,tp_k\}$  \\
    \hline
    $\mathcal{G}_R$                         &   The Resource-Linkage graph of $\mathcal{D}$, $\mathcal{G}_R=(V_R,E_R)$.    \\
    \hline
    $G^Q_R$                                 &   The Resource-Linkage graph of $Q$, $G^Q_R=\langle V^Q_R,E^Q_R \rangle$.    \\
    \hline
    $\mathcal{P}$                           &   A partition set of $\mathcal{D}$. $\mathcal{P} = \{\mathcal{P}_1, \ldots, \mathcal{P}_n\}$.   $\bigcup_{i=1}^{n} \mathcal{P}_i = \mathcal{D}_R$ .     \\
    \hline
    $\mathcal{SG}$                        &   The summary data graph that models $\mathcal{P}$. $\mathcal{SG}=\langle V_S, E_S, f_v\rangle$. \\
    \hline
    $Q^p$                                   &   A decomposition of $G^Q_R$. $Q^p=\{Q^p_1,\ldots,Q^p_m\}$. $Q^p_i=(V^p_i,E^p_i)$, $\bigcup_{i=1}^{m} E^p_i = E^Q_R$  \\
    \hline
    $\mathcal{TG}^Q$                        &   The summary query graph. $\mathcal{TG}^Q=\langle V^Q_T, E^Q_T, f^q_v\rangle$. \\
    \hline
    $\mathbb{P}$\textsubscript{\textit{1-UHC}}     &   The graph partition strategy . \\
    \hline
    \textit{Op\textsubscript{sp}}     &   The processing a sub-query on a partition. \\
    \hline
    \textit{Op\textsubscript{ipj}}     &   The  inter-partition join processing . \\
    \hline
\end{tabular}
\end{table}

\subsection{Triple Classification}
Considering that the \textit{relation} and \textit{attribute} \textit{Predicate} show different characteristics, $\mathcal{D}$ and $Q$ fall into two categories define as follow:

\begin{definition}[\textbf{R/A-Triple,R/A-Pattern}] \label{def:rdf}
A triple $t_i \in \mathcal{D}$ is called a \textit{R-Triple}(an \textit{A-Triple}) if its \textit{p} is a \textit{relation}(an \textit{attribute}). The set of all \textit{R-Triple}s(or \textit{A-Triple}s) in $\mathcal{D}$ is denoted as $\mathcal{D}_R$(or $\mathcal{D}_A$). Analogously, A triple pattern $tp_i \in Q$ is called a \textit{R-Pattern}(an \textit{A-Pattern}) if its \textit{p} is a \textit{relation}(an \textit{attribute}), and the set of \textit{R-Pattern}(\textit{A-Pattern}) in $Q$ is denoted as $Q_R$(or $Q_A$).
\end{definition}

Intuitively, a $tp_i \in Q_R$($\in Q_A$) can only be match to $t_j \in \mathcal{D}_R$($\in \mathcal{D}_A$). Using \textit{Definition}\ref{def:rdf}, we can get a compact graph model defines as follow:

\begin{definition}[\textbf{Resource-Linkage(RL) Graph}] \label{def:rlgraph}
The graph representation of $\mathcal{D}_R$(or $Q_R$) is named as the Resource-Linkage graph of $\mathcal{D}$(or $Q$), or \textit{RL graph} for brevity, denoted as $\mathcal{G}_R$ (or $G^{Q}_R$).
\end{definition}

According to \textit{Definition}\ref{def:rlgraph}, $\mathcal{G}_R$ (or $G^{Q}_R$) only models the linkage information of resources. That provides a compact representation of $\mathcal{G}$ (or $G^{Q}$). Intuitively, \textit{Property} \ref{property-rl-match} holds.

\begin{property}[\textbf{RL Graph Match}] \label{property-rl-match}
Following the \textit{subgraph match} defined in \cite{genericbacktracking}, if there exists a subgraph $g$ in $\mathcal{G}$ that match $G^Q$, consequently, the \textit{RL graph} of $g$ is a match of $G^{Q}_R$.
\end{property}

\begin{figure}
 \captionsetup{belowskip=0.5pt,aboveskip=0.5pt}
\begin{subfigure}{.48\linewidth}
  \centering
   \includegraphics[width=1.4 in]{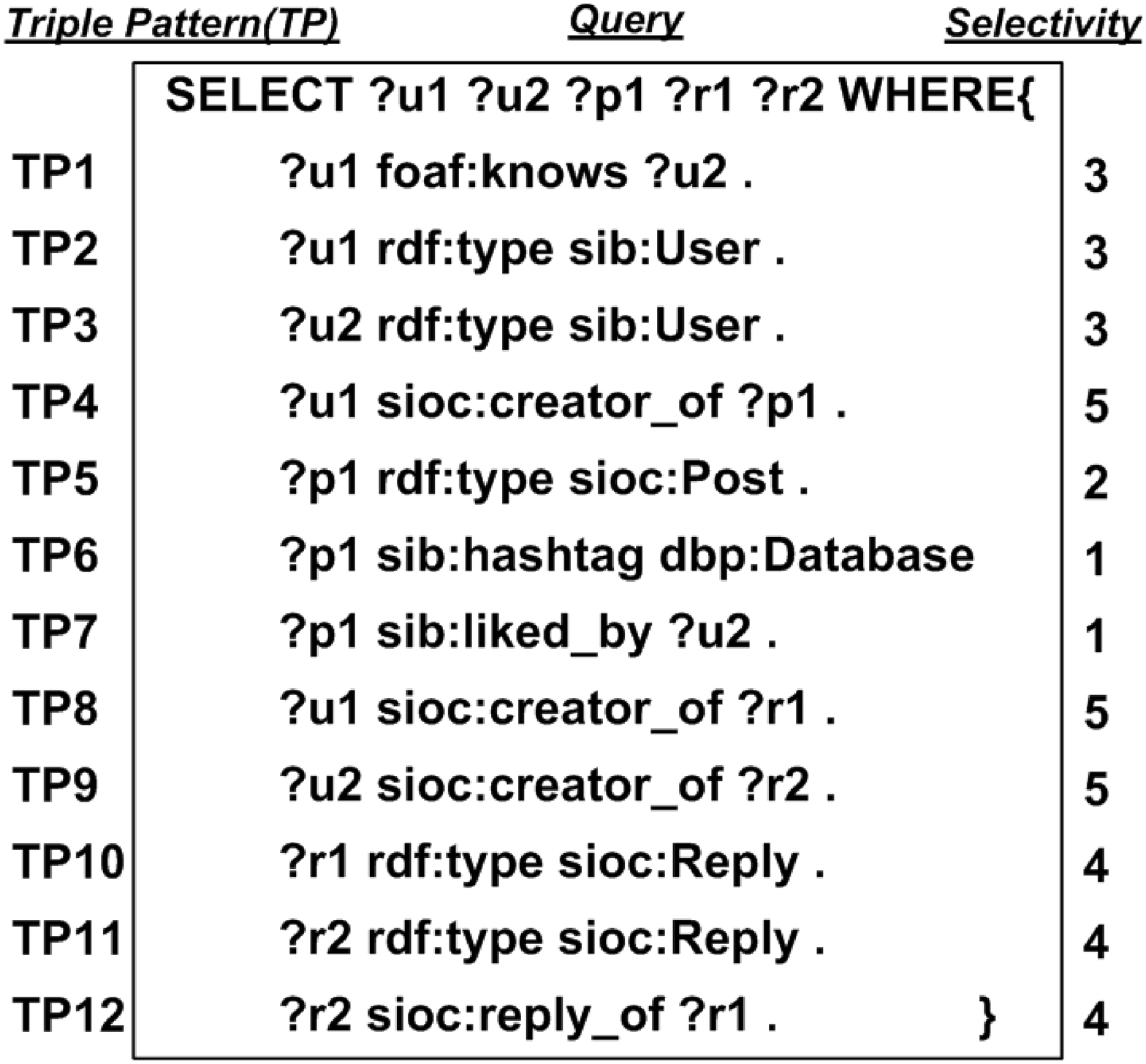}
  \subcaption{$Q$.}
  \label{subfig:query-bgp}
  \end{subfigure}
\quad
\begin{subfigure}{.43\linewidth}
  \centering
 \includegraphics[width=1.5 in]{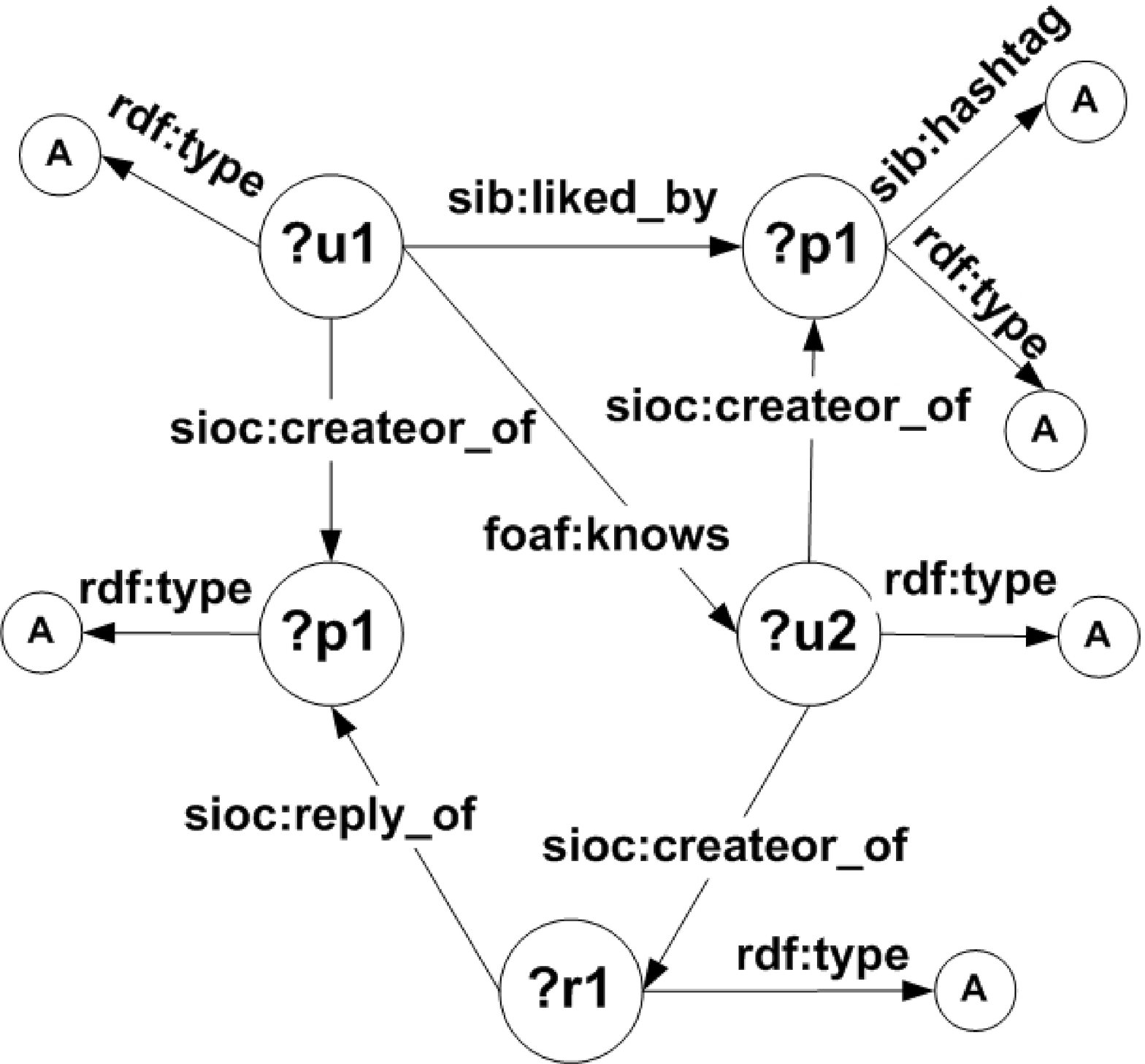}
  \subcaption{$G^{Q}$.}
  \label{subfig:query-full-graph}
  \end{subfigure}
\begin{subfigure}{\linewidth}
\centering
\includegraphics[width=3.5 in]{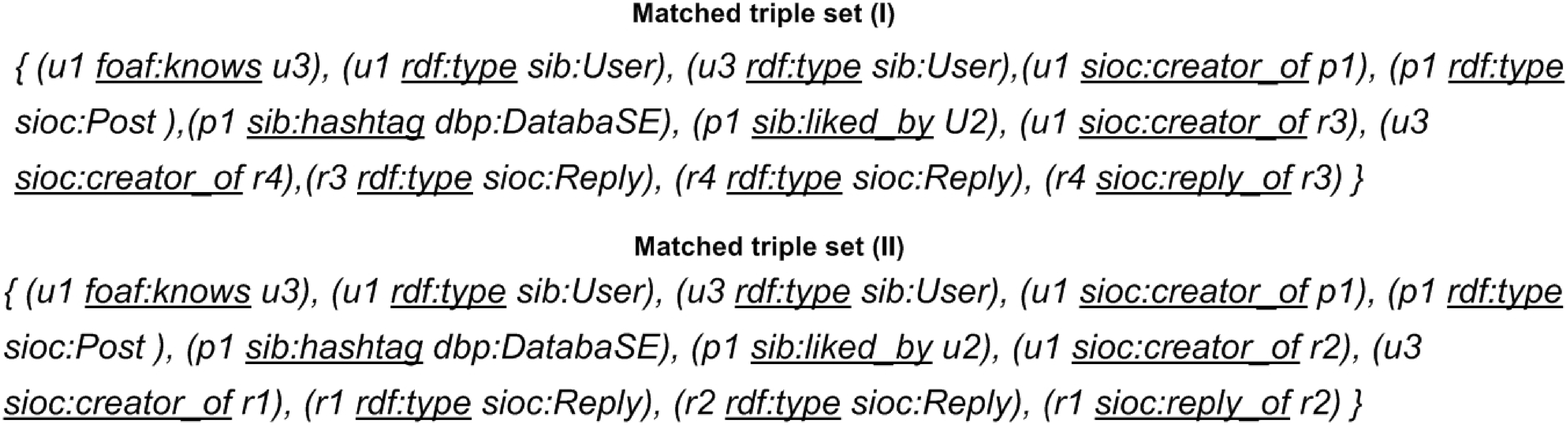}
\subcaption{Lists of matched triples of $Q$ in $\mathcal{D}$.}
\label{subfig:query-matched}
  \end{subfigure}
 \caption{An exemplar query and matched triples.}
 \label{fig:query-example}
\end{figure}

\begin{example}
\small
All triples of an RDF dataset $\mathcal{D}$  are listed in \textit{Fig.}\ref{subfig:rdf-collection}, $\mathcal{D}_A$ in the left , and $\mathcal{D}_R$ in the right.  $\mathcal{D}$ represents part of a twitter-like semantic social network, which has 3 social accounts, 2 posted contents and 4 replies. \textit{Fig.}\ref{subfig:rdf-graph} shows the equivalent \textit{RDF graph} $\mathcal{G}$, and  \textit{Fig.}\ref{subfig:rl-graph} is the RL-graph of $\mathcal{G}$. Given a query $Q$ with 12 \textit{triple patterns}(shown in \textit{Fig.}\ref{subfig:query-bgp}), its \textit{query graph} $G^Q$ is represented as \textit{Fig.}\ref{subfig:query-full-graph}. For $\mathcal{D}$, there are two matched triple sets of $Q$, lists in \textit{Fig.}\ref{subfig:query-matched}. Therefore, the result of $Q$ are \textit{\{u1,u3,p1,r3,r4\}}, and \textit{\{u1,u3,p1,r2,r1\}}.
\end{example}

\subsection{Summary Graph and Query}

For an RDF graph $\mathcal{G}$, graph partition allows triples that are close to each other in $\mathcal{G}$ to be managed in the same partition. Thus, a large $\mathcal{G}$ can be summarized into a small graph that keeps the schematic topological information among partitions. We introduce a kind of summary graph defines as follow:
\begin{definition}[\textbf{Summary RDF Graph}] \label{def:summarydatagraph}
Given an RDF dataset $\mathcal{D}$, a set $\mathcal{P} = \{\mathcal{P}_1, \ldots, \mathcal{P}_n\}$ is the partition set of its corresponding RDF graph $\mathcal{G}=\langle V,E\rangle$. A summary RDF graph $\mathcal{SG}=\langle V_S, E_S, f_v\rangle$ is a labeled, undirected multi-graph, where $V_S = \{\mathcal{P}_1, \ldots, \mathcal{P}_n\}$, and $E_S$ is consist of: \textbf{i)} edges that connects two vertexes in $V_S$ if there exists a connecting edge between two partitions; \textbf{ii)}loop edges for each vertexes in $V_S$, $e=(v,v)$,$v \in V_s$. $f_v$ maps all distinct \textit{p} in $\mathcal{P}_i$ as vertex label.
\end{definition}

Query decomposition is extensively used in converting a complex query into a set of simple sub-queries. A query $Q_R$ can be decomposed into a set of sub-queries, and each sub-query is a subgraph of the original query graph $G^Q$. We use the term \textit{subgraph} and \textit{sub-query} interchangeably in this paper. Based on query decomposition, we introduce the definition of \textit{transformed query graph} as follow:

\begin{definition}[\textbf{Transformed Query Graph}] \label{def:summaryquerygraph}
Given a query $Q$ and its corresponding query graph $G^Q=\langle V^Q,E^Q\rangle$. Let one of its decomposition as $Q^p=\{Q^p_1,\ldots,Q^p_m\}$, where $Q^p_i=\langle V^p_i,E^p_i\rangle$, $\bigcup_{i=1}^{m} E^p_i = E^Q$. A \textit{transformed query graph} $\mathcal{TG}^Q=\langle V^Q_T, E^Q_T, f^q_v\rangle$ for $Q$ is a vertex-labeled, undirected graph, where $V^Q_T = \{Q^p_1,\ldots,Q^p_m\}$, and edge $(Q^p_i,Q^p_j) \in E^Q_T$ if their exists common vertices between $Q^p_i$ and $Q^p_j$. $f^q_v$ maps all distinct \textit{p} in $Q^p_i$ as vertex label.
\end{definition}
\begin{example}
\small
We partition the $\mathcal{G}_R$ in \textit{Fig.} \ref{subfig:rl-graph} into two partitions, $\mathcal{P}^o = \{\mathcal{P}^o_1, \mathcal{P}^o_2\}$, show as \textit{Fig.} \ref{subfig:graph-partition}. Using $\mathcal{P}^o$, we can model a \textit{Summary RDF Graph} $\mathcal{SG}$ as \textit{Fig.} \ref{subfig:summary-graph}. For the query in \textit{Fig.} \ref{subfig:query-bgp}, its query \textit{RL-Graph} $Q_R$ can be decomposed into four subgraphs $Q^p=\{Q^p_1,\ldots,Q^p_4\}$, show as \textit{Fig.} \ref{subfig:query-rewrite}. Using $Q^p$, we model a \textit{transformed query graph} $\mathcal{TG}^Q$ as \textit{Fig.} \ref{subfig:transformed-graph}.
\end{example}

\subsection{Summary Graph Match}

Lets \textit{OP\textsubscript{sp}} refers to the operation of a sub-query on a data partition. More precisely, the processing of a $Q^p_i \in Q^p$ on a $\mathcal{P}_a \in \mathcal{P}$ is denoted as $Q^p_i(\mathcal{P}_a)$, and its result set is denoted as $R(Q^p_i(\mathcal{P}_a))$. To deal with the connection between sub-queries, we introduce the concept of intra-partition join, expressed as \textit{OP\textsubscript{ipj}}. For two \textit{OP\textsubscript{sp}}， namely $Q^p_i(\mathcal{P}_a)$ and $Q^p_i(\mathcal{P}_a)$, $1\leq i,j\leq m$, $1\leq a,b\leq n$, a \textit{OP\textsubscript{ipj}} is the join processing between $R(Q^p_i(\mathcal{P}_a))$ and $R(Q^p_j(\mathcal{P}_b))$, denoted as \textit{OP\textsubscript{ipj}}$(Q^p_i(\mathcal{P}_a),Q^p_j(\mathcal{P}_b))$, and the result set is denoted as $ R(Q^p_i(\mathcal{P}_a),Q^p_j(\mathcal{P}_b))$. We define the summary graph match of $\mathcal{TG}^Q$ on $\mathcal{SG}$ as follow:

\begin{definition}[\textbf{Summary Graph Match}] \label{def:match}
Given a $\mathcal{TG}^Q=\langle V^Q_T, E^Q_T, f^q_v\rangle$ and a $\mathcal{SG}=\langle V_S, E_S, f_v\rangle$, where $V^Q_S=\{Q^p_1,\ldots,Q^p_m\}$,$V_S= \{\mathcal{P}_1, \ldots, \mathcal{P}_n\}$, a \textit{summary graph match} is an injective function $M : V^Q_T\rightarrow V_S$, such that: \textbf{i) vertex match:} $\forall Q^p_i \in V^Q_T$, $f^q_v(Q^p_i) \subseteq f_v(M(Q^p_i))$, and  $R(Q^p_i$( $M(Q^p_i))) \neq \varnothing$, and
\textbf{ii) edge injection:} $\forall(Q^p_i,Q^p_j) \in E^Q_T$, $(M(Q^p_i),M(Q^p_j)) \in E_S$, and $R(Q^p_i(M(Q^p_i)), Q^p_j(M(Q^p_j))) \neq \varnothing$.
\end{definition}

%
%
%
%

\section{Overview of SGDQ} \label{sec:overview}

\begin{figure}
  \centering
  \includegraphics[height=0.25\textheight]{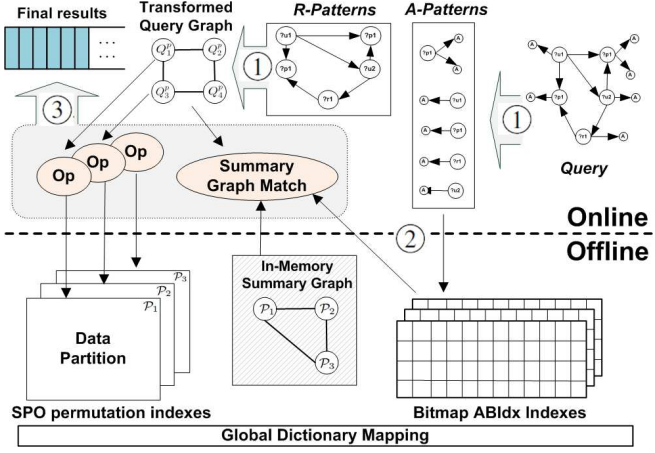}
\caption{The Architecutre of SGDQ framework.}
\label{fig:paradigm}
\end{figure}

\textit{Fig.} \ref{fig:paradigm} gives an overview of SGDQ framework. A dataset $\mathcal{D}$ are offline indexed to facilitate online query processing.

\subsection{Offline Processing}

Given an RDF dataset $\mathcal{D}$. according to the \textit{p} of each triple, $\mathcal{D}$ is divided into two parts, $\mathcal{D}_A$ and $\mathcal{D}_R$. $\mathcal{D}_A$ is managed on disk using a set of bitmap-based indexes(\textit{Section} \ref{subsec-bitmapidx}). Besides, $\mathcal{D}_R$ is modeled as a \textit{data RL-graph} $\mathcal{G}_R$, and partitioned into a set of subgraphs $\mathcal{P}^{o}=\{\mathcal{P}^{o}_{1}, \ldots ,\mathcal{P}^{o}_{n}\}$ using METIS algorithm\cite{Karypis95metis}(denoted as $\mathbb{P}$\textsubscript{\textit{METIS}}). We call $\mathcal{P}^{o}$ as an \textit{original partitions}. $\mathbb{P}$\textsubscript{\textit{METIS}} is an edge-cut graph partition method, it divides a graph into a given number(there it is $n$) of roughly vertices-balanced connected subgraphs $\mathcal{P}^{o}_{j} =(V^o_j,E^o_j),1\leq j \leq n$, with minimum number of cutting edges. Therefore, we can model $\mathcal{P}^{o}$ as a summary data graph $\mathcal{SG}$ according to \textit{Definition} \ref{def:summarydatagraph}. $\mathcal{SG}$ is maintained as an in-memory adjacency list to facilitate graph exploration. Each partition in $\mathcal{P}^o$ is further expanded using a partition strategy which duplicates boundary triples, and final partitions $\mathcal{P} = \{\mathcal{P}_1,\ldots,\mathcal{P}_n \}$ are generated. Each $\mathcal{P}_i \in \mathcal{P}$ is managed as a self-governed triple store. All triples in $\mathcal{D}_R$ are encoded globally using a global dictionary. This enables inter-partition processing. (\textit{Section} \ref{subsec:1-uhc}).

\subsection{Runtime Query Processing} \label{subsec:runtime}

Once a user interactively submit a query $Q$, the query processing is implemented in three steps.

\noindent \textit{Step} \circled{1}, \textit{Query preprocessing}. Under the premise of no \textit{Predicate} join in $Q$ , $Q$ is divided into $Q_A$ and $Q_R$, judged by whether \textit{p} of each triple pattern is a \textit{relation} or an \textit{attribute}. $Q_R$ is modeled as a $\mathcal{G}^Q_R$, while $Q_A$ are grouped by \textit{Subject}. By decomposing $\mathcal{G}^Q_R$ into set of sub-queries $Q^p=\{Q^p_1,\ldots,Q^p_m\}$, we model a $\mathcal{TG}^Q$ (\textit{Section} \ref{subsec:querydecomp}).

\noindent \textit{Step} \circled{2}, \textit{A-Pattern processing}. Foremost, each group of $Q_A$ is processed using bitmap indexes, and the candidate set of the corresponding vertex in $\mathcal{G}^Q_R$ are generated for further used in \textit{OP\textsubscript{sp}}.(\textit{Section} \ref{subsec-bitmapidx}).

\noindent \textit{Step} \circled{3}, \textit{Summary graph driven processing}. Based on data partition and query decomposition, the query processing is decomposed as a set of light-weight \textit{OP\textsubscript{sp}}. Follow the paradigm of match $\mathcal{TG}^Q$ on $\mathcal{SG}$, We organize the execution of \textit{OP\textsubscript{sp}} and perform \textit{OP\textsubscript{ipj}} between two \textit{OP\textsubscript{sp}}. The final results is generated in an incremental fashion, and stops until no more matches could be found. (\textit{Section} \ref{subsec:sgdq-algo}).

%
%
%
%
\section{Data Organization and Operations}\label{sec-partition}

In this section, we explain how to decompose a costly query processing a series of light-weight \textit{Op\textsubscript{sp}} operations. A common practice is to introduce data partition and query reformulation. Our implementation is efficient in two aspects. On one side, we identify two types of query pattern to simplify a query. Utilize the characteristics of \textit{A-Pattern}, we introduce bitmap index for \textit{A-Triples} and implement efficient bitwise operation for \textit{A-Pattern} processing(\textit{Section} \ref{subsec-bitmapidx}). On the other side, we introduce a data partition strategy(\textit{Section} \ref{subsec:1-uhc}) and corresponding query decomposition method, and implement an efficient \textit{Op\textsubscript{sp}}(\textit{Section} \ref{subsec:querydecomp}).

\begin{figure}
\captionsetup{belowskip=0.5pt,aboveskip=0.5pt}
\captionsetup[sub]{margin=0.5pt,skip=0.5pt, labelfont=bf}
\begin{subfigure}{.43\linewidth}
  \centering
  \includegraphics[height= 1.5 in, width=1.2 in]{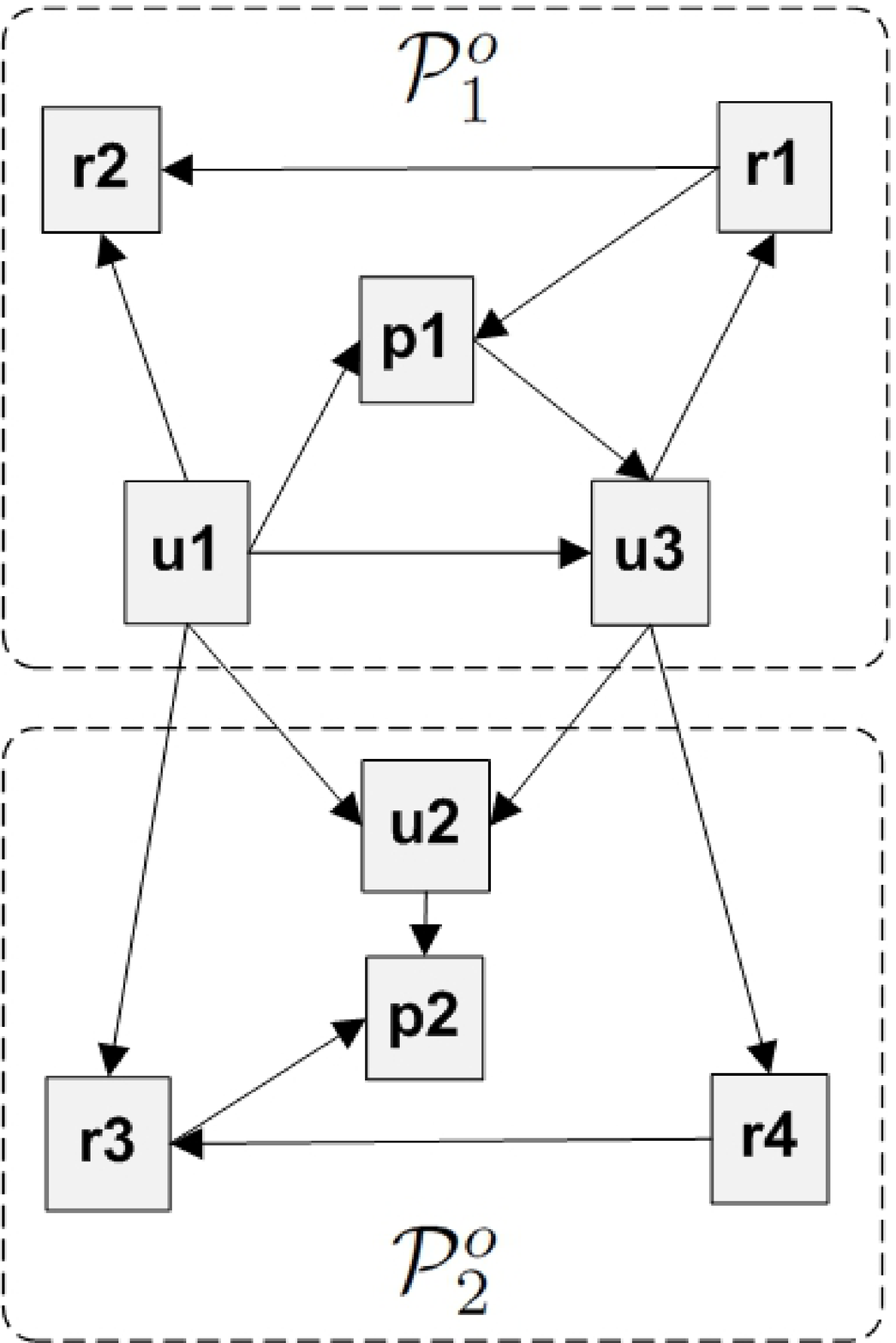}
  \subcaption{The original partition, \qquad \qquad  $\mathcal{P} = \{\mathcal{P}^0_1, \mathcal{P}^0_2\}$.}
  \label{subfig:graph-partition}
\end{subfigure}
\begin{subfigure}{.52\linewidth}
  \centering
  \includegraphics[height= 1.6 in, width=1.6 in]{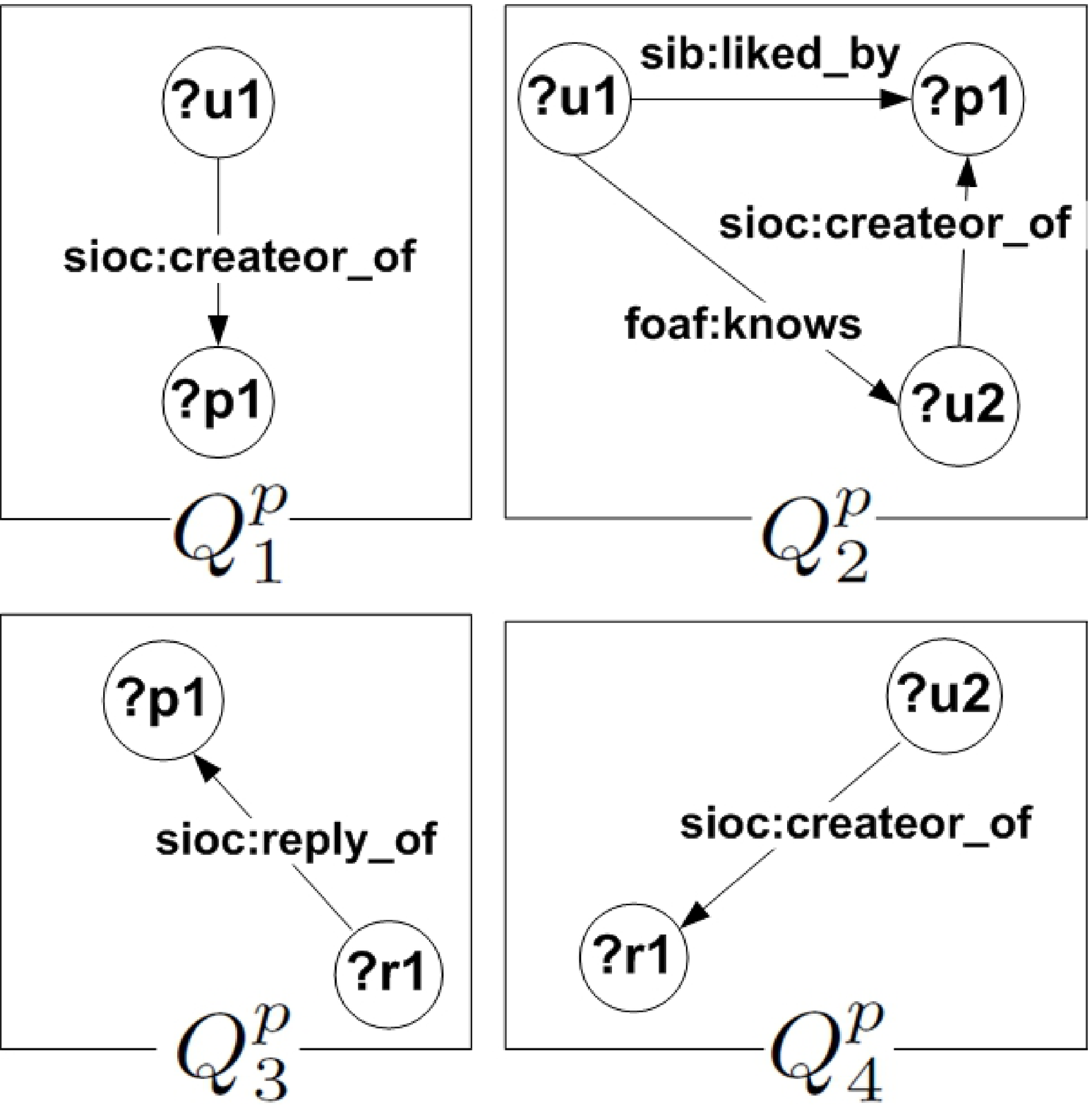}
  \subcaption{$Q^p = \{Q^p_1, Q^p_2, Q^p_3, Q^p_4\}$.}
  \label{subfig:query-rewrite}
\end{subfigure}
  \vspace{\baselineskip}
\begin{subfigure}{.45\linewidth}
  \centering
  \vspace{\baselineskip}
  \vspace{\baselineskip}
  \includegraphics[height= 0.6 in,width=1.5 in]{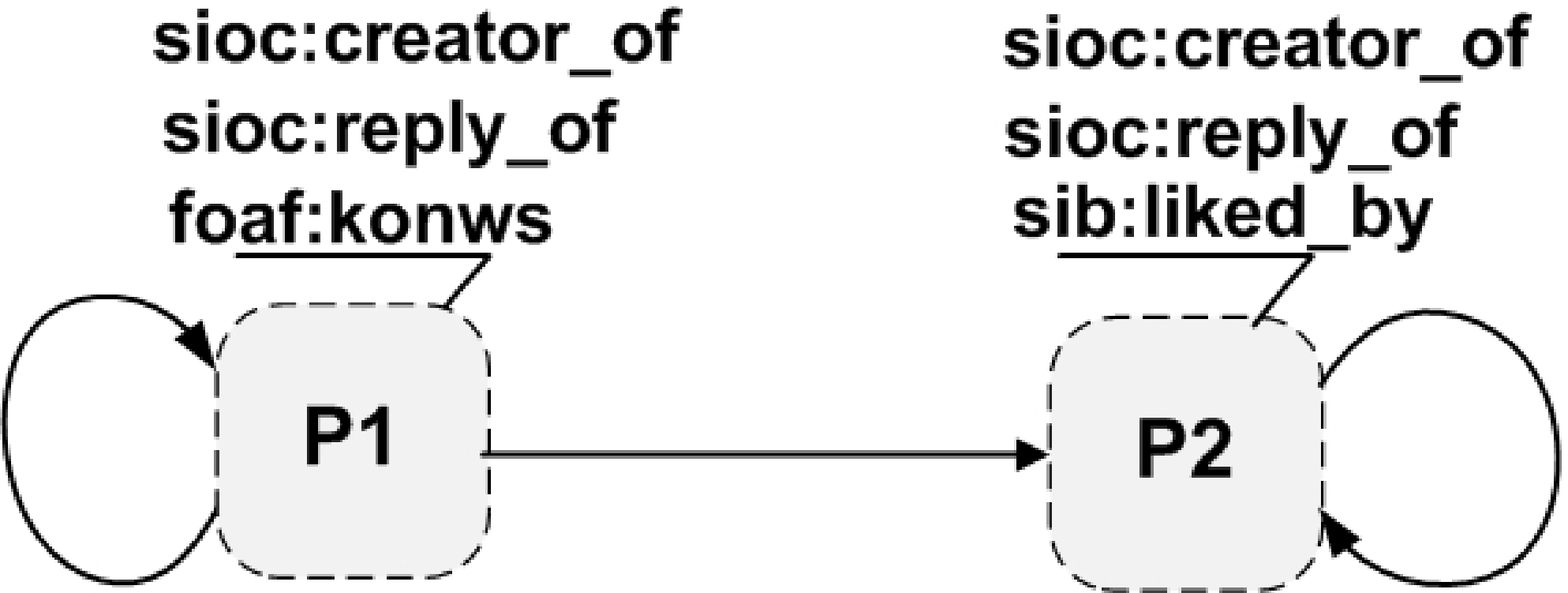}
  \vspace{\baselineskip}
  \subcaption{Summary graph, $\mathcal{SG}$.}
  \label{subfig:summary-graph}
\end{subfigure}
\begin{subfigure}{.52\linewidth}
  \centering
  \includegraphics[height= 1.0 in, width=1.2 in]{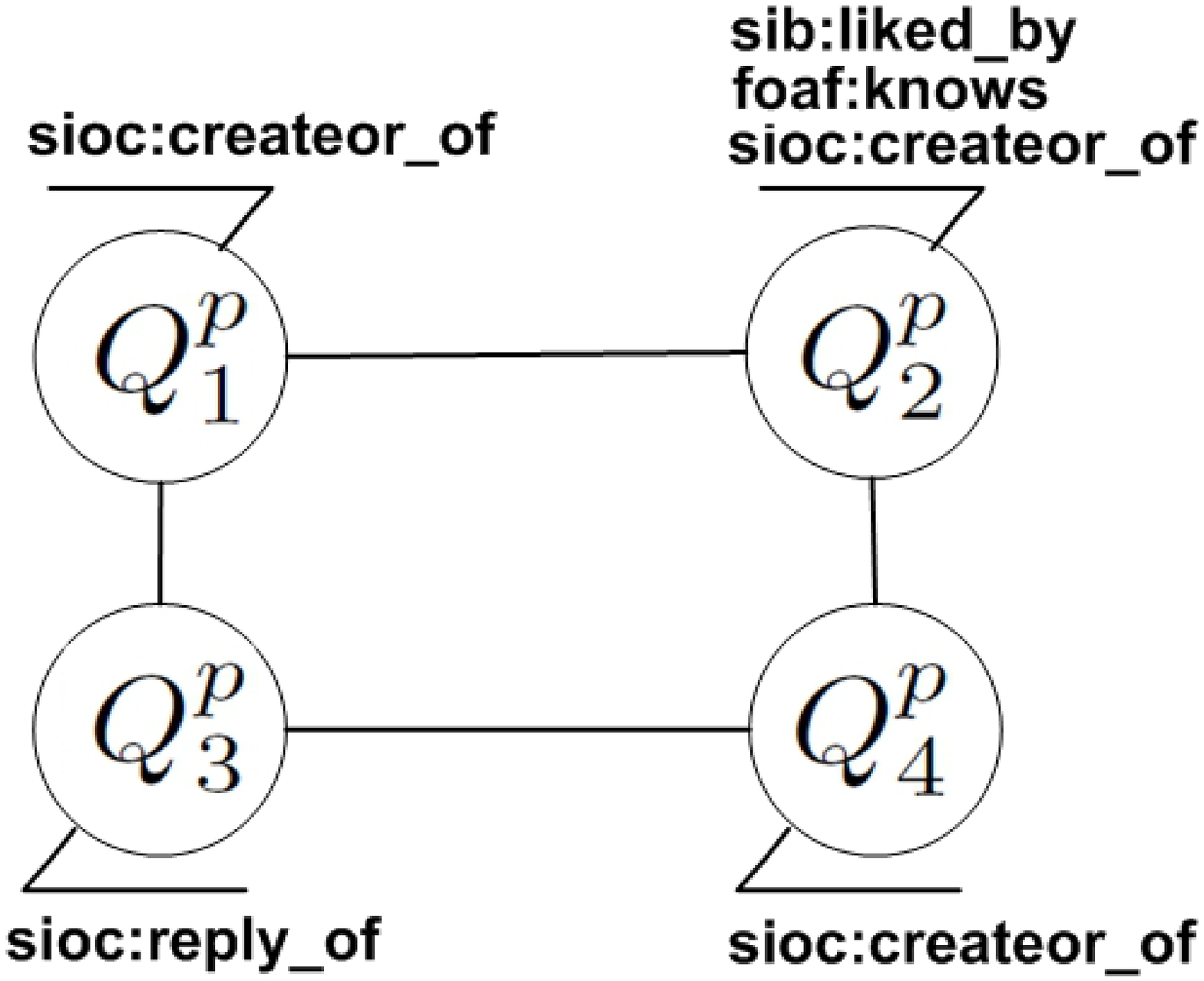}
  \subcaption{Transformed query graph, $\mathcal{SG}^Q$.}
  \label{subfig:transformed-graph}
\end{subfigure}
\caption{An example of data partition and query rewrite.}
\end{figure}

\subsection{Bitmap Indexes for \textit{A-Triple}s and \textit{A-Pattern} processing} \label{subsec-bitmapidx}

Managing and querying \textit{A-Triple}s in a specific manner brings two main benefits. First, is avoids unnecessary duplication of \textit{A-Triple}s in data partitions. This simplifies data partition. Second, it employs efficient bitmap-based bitwise operations that take advantage of the characteristics of \textit{s-s join} for \textit{A-Pattern}s.

Bitmaps are commonly used in databases as they can be efficiently compressed and processed. Using bitmap in RDF data, \textit{s-s joins} on a common \textit{Subject} can be translate to simple bitwise AND operations. This saves a potentially large number of joins compared with canonical relational query processing. For this reason, some RDF researches\cite{bitmat-www2010,triplebit-vldb2013} have also considered the extensive adoption of bitmap indexes. To encode data as bitmap, we build two types of global dictionary, one for $\mathcal{D}_R$ and the other for $\mathcal{D}_A$, denoted as $\mathcal{M}_R$ and $\mathcal{M}_A$ respectively. $\mathcal{M}_R$ holds the mapping of an unique ID, namely \textit{VertexID}, to a \textit{s} or an \textit{o} in $\mathcal{D}_R$(i.e. $v \in V_R$). $\mathcal{M}_R$ is built according to the \textit{original partitions}. We consecutively allocate a \textit{VertexID} to each vertex grouped by partitions, and store all \textit{VertexID}s in an sorted order in $\mathcal{M}_R$. Similarly,  $\mathcal{M}_A$ is the mapping of an  \textit{AttributeID} to an \textit{o} in $\mathcal{D}_A$. All \textit{AttributeID}s are managed as a sorted  list in $\mathcal{M}_A$. The partial order lists of \textit{VertexID} and \textit{AttributeID} in $\mathcal{M}_R$ and $\mathcal{M}_A$ are further used in bitmap indexes construction.

Based on encoding data globally using $\mathcal{M}_R$ and $\mathcal{M}_A$, we design and implement a set of compressed column-oriented indexes that dedicated for \textit{A-Triples} management and \textit{A-Pattern}s processing. Consider there are two types of \textit{A-Pattern},
\begin{description}
\item[\textit{Pattern-I}] \ \textit{(?s,p,o)}, where \textit{o} is a designated URI or Literal.
\item[\textit{Pattern-II}] \ \textit{(?s,p,?o)} and \textit{(s,p,?o)}, where \textit{?o} is a variable.
\end{description}

For \textit{Pattern-I}, we design a bitmap of size $|V_R|$, referred to as \textit{BitVector}, to represent the existence of all \textit{VertexID} for a given combination of \textit{o} and \textit{p}. In a \textit{BitVector}, a bit at position $i$ corresponds to the $i$-th \textit{VertexID} in $\mathcal{M}_R$, and is set to 1 if triple (\textit{VertexID},\textit{p},\textit{o}) $\in \mathcal{D}_A$, or 0 otherwise. There are totally $|\mathcal{D}_A|$ such \textit{BitVector}s. They are grouped by \textit{p}, and each group is sorted and indexed on \textit{o} using B+-tree to facilitate index lookup. Consider that the size of \textit{BitVector}s (i.e. $|V_R|$) can be quite large, a common way in practice is using compression to take advantage of the sparse nature of the vectors. Similar to \cite{bitmat-www2010}, we adopts D-Gap compression\footnote{\url{http://bmagic.sourceforge.net/dGap.html}}. This allows bitwise operations to be performed against compressed data, which eliminates the cost of \textit{AttributeVector} decompression. Such index can be viewed as a special implementation of \textbf{POS} permutation index. We name it as \textsf{ABIdx}\textsubscript{POS}.

For \textit{Pattern-II}, we design two kinds of index, namely, \textsf{ABIdx}\textsubscript{comp} and \textsf{ABIdx}\textsubscript{PSO}. In \textsf{ABIdx}\textsubscript{comp}, each \textit{p} maintains a bitmap which stores the bitwise AND results of all \textit{BitVector}s of this \textit{p} in \textsf{ABIdx}\textsubscript{POS}. This bitmap captures all \textit{A-Triples} that have this \textit{p}.  \textsf{ABIdx}\textsubscript{PSO} can be viewed as a special implementation of \textbf{PSO} index, and implemented the same way as \textsf{ABIdx}\textsubscript{POS}, where a \textit{BitVector} of size $|\mathcal{M}_A|$ is arranged according to the sorted list of \textit{AttributeID} in $\mathcal{M}_A$.

At query time, for a $v \in V^Q_R$ of a given $Q$, let $TP_A$ be the set of \textit{A-Pattern}s that have $v$ as \textit{Subject}. We introduce an operation, named \textit{CandidateRLVertex}, that inputs $TP_A$ and returns the existence of candidate \textit{VertexID}s for $v$ as a \textit{BitVector} of size $|V_R|$. For each pattern $tp_i \in TP_A$, \textit{CandidateRLVertex} first check the type of $tp_i$. If $tp_i$ a \textit{Pattern-I}, it retrieves \textsf{ABIdx}\textsubscript{PSO} according to \textit{p} and \textit{o} of $tp_i$ and get a \textit{BitVector}. Otherwise, if $tp_i$ a \textit{Pattern-II}, it gets a \textit{BitVector} in \textsf{ABIdx}\textsubscript{comp} corresponding to the \textit{p} of $tp_i$. The final result is a  \textit{BitVector} that is derived from the bitwise AND of all retrieved \textit{BitVector}s. This \textit{BitVector} represents the existence of all \textit{VertexID}s. To get the corresponding \textit{AttributeID} for the variable \textit{?o} in \textit{Pattern-II}, we implement an extra operation, named \textit{RetrieveAttribute}, at the final stage of $Q$ processing. \textit{RetrieveAttribute} retrieves the \textit{BitVector} in \textsf{ABIdx}\textsubscript{POS} according to the \textit{VertexID} in previous results and \textit{p} in \textit{Pattern-II}, then retrieves $\mathcal{M}_A$ to get the corresponding \textit{AttributeID}, and maps to its real value.

\begin{example}
\small
For the \textit{original partition} (\textit{Fig.}\ref{subfig:graph-partition}) of $\mathcal{D}$ ( \textit{Fig.}\ref{subfig:rdf-collection}), we arrange an ordered vertex list as \{u1,u3,p1,r1,r2\} for $\mathcal{P}^o_1$ and \{u2,p2,r3,r4\} for $\mathcal{P}^o_1$. In $\mathcal{M}_R$, we consecutively allocate \textit{VertexID} 1 to 5 for the former list, and 6 to 9 for the later one. Given a query with 2 patterns, $Q$=\{\textit{?s rdf:type sioc:Reply}, \textit{?s sioc:content ?o}\}, we first convert all strings into ID using $\mathcal{M}_A$ and $\mathcal{M}_R$, then retrieve \textsf{ABIdx}\textsubscript{PSO} according to \textit{Predicate} \textit{rdf:type} and \textit{Object} \textit{sioc:Reply}, and get a \textit{BitVector} of [000110011], Meanwhile,\textsf{ABIdx}\textsubscript{comp} is retrieved using a key of \textit{Predicate} \textit{sioc:content}, and [001110111] is got. By performing bitwise AND between two retrieved \textit{BitVector}, we get a result \textit{BitVector} [000110011]. To get the final results, we get all 1-bit corresponding \textit{VertexID}s using $\mathcal{M}_R$, and retrieve \textsf{ABIdx}\textsubscript{POS} for each  \textit{VertexID} and \textit{Predicate} \textit{sioc:content}. and get a result \textit{BitVector} each time. The final result can be derived by checking the 1-bit corresponding \textit{AttributeID} in $\mathcal{M}_A$.
\end{example}

\subsection{\textit{R-Triple}s Partition and Index}\label{subsec:1-uhc}

Next, we introduce how $\mathcal{D}_R$ is partitioned. In order to explain what type of query $Q$ can be implemented without inter-partition operations, we introduce the concept of \textit{query coverage}. Given a set of partitions $\mathcal{P}=\{\mathcal{P}_{1}, \ldots, \mathcal{P}_{n}\}$ generated by a partition strategy $\mathbb{P}$ over $\mathcal{D}$. If a query $Q$ can be completely answered in $\mathcal{P}$ without inter-partition operations, we call that partition $\mathcal{P}$ can \textit{cover} $Q$ .If $\mathbb{P}$ can completely \textit{cover} all query graphs in which the shortest path between two arbitrary vertices $ < \phi$, we call $\phi$ as the \textit{query coverage} of $\mathbb{P}$. To increase $\phi$, a common way is to introduce triple replication in each $\mathcal{P}_{i}$, $1\leq i \leq n$. We defines \textit{replication factor} $\alpha = \frac{|\mathcal{D}_A| + \sum^{n}_{i=1}|\mathcal{P}_{i}| }{|\mathcal{D}|}$ to quantitatively measure such replication. There is a tradeoff between $\phi$ and $\alpha$ . For example, $\mathbb{P}$\textsubscript{\textit{METIS}} have $\phi=0$, $\alpha=1$. This means although $\mathbb{P}$\textsubscript{\textit{METIS}} does not introduces any replicated triples, it must deal with inter-partition query processing for an arbitrary  query. Increase $\phi$ can results in an exponential increase of replicated triples, which led to an extra storage overhead.

\begin{figure}
  \centering
 \captionsetup{belowskip=0pt,aboveskip=0.5pt}
 \captionsetup[sub]{belowskip=0pt,aboveskip=0pt, margin=0pt,skip=0pt, labelfont=bf}
 \begin{minipage}[b]{0.45\linewidth}
  \centering
   \includegraphics[width=1.3 in]{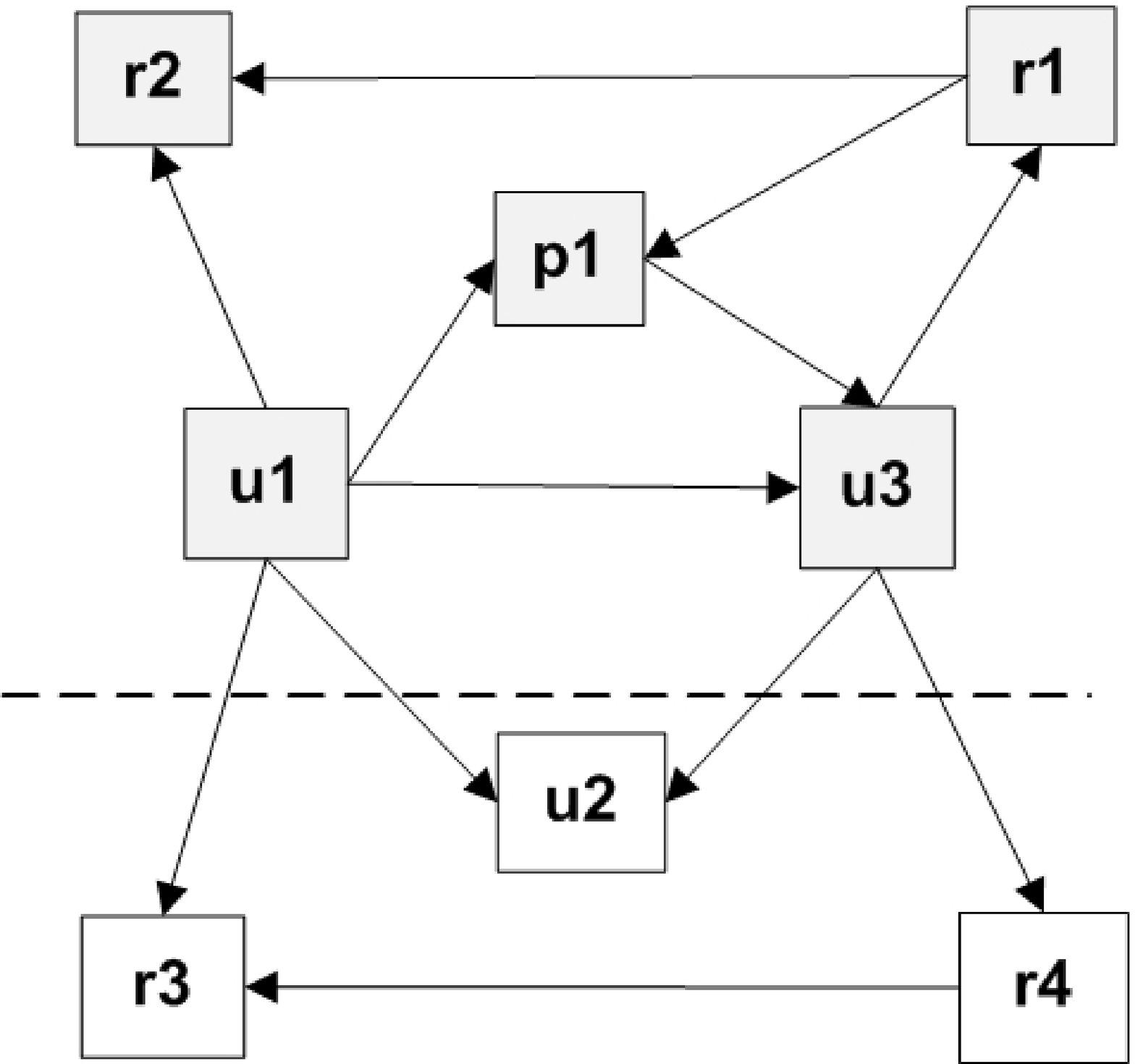}
  \vspace{\baselineskip}
  \subcaption{$\mathcal{P}_1$.}
  \label{subfig:partition-original}
\end{minipage}%
\quad
\begin{minipage}[b]{0.45\linewidth}
  \centering
 \raisebox{0.06\height}
 {\includegraphics[width=1.3 in]{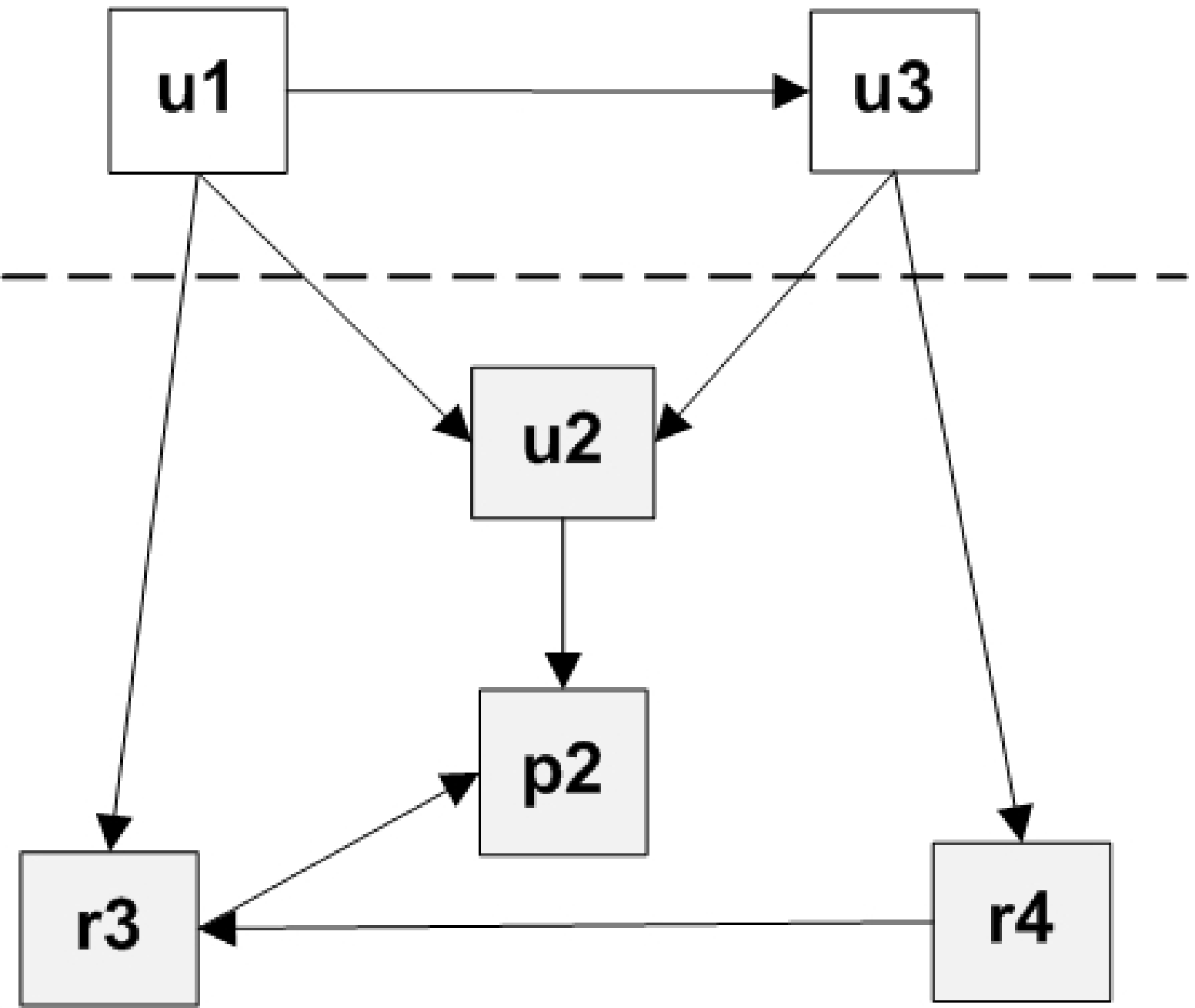}}
  \vspace{\baselineskip}
  \subcaption{$\mathcal{P}_2$.}
  \label{subfig:partition-1ohc}
\end{minipage}%
 \caption{A raw partition and its 1-OHC partitions.}
 \setlength{\belowcaptionskip}{-16pt}
 \label{fig:partition}
\end{figure}

To achieve a reasonable $\phi$ with acceptable $\alpha$, in our work, we utilize a $\mathbb{P}$ named \textit{Undirected One Hop Cover}(1-UHC) partition strategy, denoted as $\mathbb{P}$\textsubscript{\textit{1-UHC}}. $\mathbb{P}$\textsubscript{\textit{1-UHC}} follows the idea of \textit{undirected 1-hop guarantee} introduced by \cite{1UHC}. To elaborate, based an \textit{original partition} $\mathcal{P}^{o}_{j} =(V^o_j,E^o_j) \in \mathcal{P}^{o}$ for $\mathcal{G}_R$, let $V_c = \{v | (v,v_i) \in E^o_j \vee (v_i,v) \in E^o_j, v_i \in V^o_j, v \notin V^o_j \}$ denotes the set of cutting edges in $\mathcal{P}^{o}_{j}$. $\mathbb{P}$\textsubscript{\textit{1-UHC}} generates expanded partitions $\mathcal{P}_{j}=(V_j,E_j)$ following these steps:
\begin{enumerate}
\item Adds all cutting edges $ \forall e \in \{e=(v_a,v_b) \in \mathcal{G}_R |  (v_a \in V^o_j \wedge v_b \in V_c ) \vee (v_a \in V_c \wedge v_b \in V^o_j)\}$ to $E_j$;
\item Adds all edges that connect cutting edges  $ \forall e \in \{ e=(v_a,v_b) \in \mathcal{G}_R | (v_a \in V_c \wedge v_b \in V_c) \} $ to $E_j$;
\item Assigns $\forall v \in V_c$ to $V_j$.
\end{enumerate}
$\mathbb{P}$\textsubscript{\textit{1-UHC}} differs from \cite{1UHC} in that it solely applied on $\mathcal{G}_{R}$. As $\mathcal{G}_R$ is more compact than $\mathcal{G}$, this leds to a smaller $\alpha$.
\begin{example}
\small
For $\mathcal{G}_R$ in \textit{Fig.}\ref{subfig:rl-graph}, an \textit{original partition} $\mathcal{P}^{o}=\{\mathcal{P}^{o}_{1} ,\mathcal{P}^{o}_{2}\}$ generated by  $\mathbb{P}$\textsubscript{\textit{METIS}} are shown as \textit{Fig.}\ref{subfig:partition-original}. After apply $\mathbb{P}$\textsubscript{\textit{1-UHC}}, we get partition $\mathcal{P}=\{\mathcal{P}_{1} , \mathcal{P}_{2}\}$  shown as  \textit{Fig.}\ref{fig:partition}.
\end{example}

For management of triples in a partition, we independently index and organize them as a set of permutation indexes. This is supported by most state-of-the-art triple store. There we adopted RDF-3X\cite{rdf3x2010}. To enable joins between partitions, for each partition, we use the \textit{VertexID} in $\mathcal{M}_R$ to globally encode the \textit{s} and \textit{o} of all triples. We additionally maintain two metadata of a partition. One represents all \textit{VertexID} in a $\mathbb{P}$\textsubscript{\textit{1-UHC}} partition. It is stored as a \textit{BitVector}, named as \textit{pVector}. \textit{pVector} encodes the existence of $v_i \in V_R$ in a partition follows the partial order of \textit{VertexID} in $\mathcal{M}_R$. It is useful in filtering triples using \textit{A-Pattern}s. The other represents all \textit{VertexID}s in an \textit{original partition}. As they are allocated consecutively in $\mathcal{M}_R$, we only need to store the start and end \textit{VertexID}. This is useful in avoiding results duplication. We discuss the use of indexes and metadata in the subsequent \textit{Section} \ref{subsec:querydecomp}.

\noindent \textbf{Selection of \textit{n}} is an empirical problem. A premise of our work is that each partitions can be efficiently processed without too much memory thrashing. This led to a rough estimation that the size of a partition and the possible intermediate results for arbitrary query processing can be a well fit to the restricted memory. In our evaluation, we adopt $n=500$.

\subsection{Query decomposition and Sub-query Processing}\label{subsec:querydecomp}

Next, we explain how to decompose $G^Q_R$ into a set of subgraphs $Q^p=\{Q^p_1,\ldots,Q^p_m\}$ according to $\mathbb{P}$\textsubscript{\textit{1-UHC}}, and the light-weight implementation of \textit{OP\textsubscript{sp}} considering \textit{A-Pattern}s. Furthermore, we introduce a mechanism in \textit{OP\textsubscript{sp}} to avoid generating duplications in final query results.

According to \textit{Definition} \ref{def:summaryquerygraph}, a \textit{transformed query graph} $\mathcal{TG}^Q$ is constructed by transform each $Q^p_i$ into a $\mathcal{TG}^Q$ vertex, and establish an edge between two $\mathcal{TG}^Q$ vertices if their corresponding subgraphs in $Q^p$ share a common vertex in $Q^G_R$. As each $Q^p_i \in Q^p$ can by covered by $\mathbb{P}$\textsubscript{\textit{1-UHC}}, and $\mathbb{P}$\textsubscript{\textit{1-UHC}} has $\phi=2$, there are  only two types of sub-queries :
\begin{description}
\item[\textit{Type-I}] \textit{simplex triple pattern}: a single pattern in $G^Q_R$.
\item[\textit{Type-II}] \textit{complete graph pattern}: a set of triple patterns that forms an undirected complete graph in $G^Q_R$. The most simple case is a triangle of three patterns.
\end{description}

The generation of a transformed query graph can be implemented in a greedy fashion. We start with selection of a vertex $v$ with the maximum degree in $G^Q_R=\langle V^Q_R,E^Q_R \rangle$, then traverse through all $v$'s neighbour vertices to check whether a complete graph exists. If exists, we add a vertex to $\mathcal{TG}^Q$, and add an edge between existing vertices in $\mathcal{TG}^Q$ to newly added vertex if they share a common vertex. If all complete graph in $G^Q_R$ that contains $v$ has been detected, we  remove current vertex $v$ from $V^Q_R$ and all edges of all complete graph from $E^Q_R$. We iterate the above process until there are no edges in $E^Q_R$. Obviously in worst case, $\mathcal{TG}^Q$ can have $|E^Q_R|$ vertices.

As the basic building block of SGDQ, an \textit{OP\textsubscript{sp}}, $Q^p_i(\mathcal{P}_j)$, is the sub-query $Q^p_i \in Q^p, 1\leq i\leq m$ on partition $\mathcal{P}_j \in \mathcal{P}, 1\leq j\leq n$. Optimization of \textit{OP\textsubscript{sp}} can greatly enhance the overall query performance. Furthermore, as $\mathbb{P}$\textsubscript{\textit{1-UHC}} introduce overlapping among partitions, in \textit{OP\textsubscript{sp}} a mechanism is needed to tackle the problem of latent duplications in final results.

Before explain the implementation of \textit{OP\textsubscript{sp}}, we first introduce a process, named as \textit{filter}. The \textit{filter} uses \textit{A-Pattern}s to prune the matched triple set of a \textit{R-Pattern} in a $\mathcal{P}_j$. Given a \textit{R-Pattern}, let $v_o$ be its \textit{Subject} vertex, $v_s$ be its \textit{Object} vertex. Each vertex may have \textit{A-Pattern}s connect to it, and the candidate set of $v_s$ and $v_o$ , denoted as $CR(v_s)$ and $CR(v_o)$, can be got using \textit{CandidateRLVertex} operation. For all triples in $\mathcal{P}_j$ that match a \textit{R-Pattern}, \textit{filter} eliminates triples that theirs \textit{Subject}(or \textit{Object}) are not contained in $CR(v_o)$(or $CR(v_s)$). The set containment check is also based on \textit{BitVector}s, and can be efficiently implemented in $O(1)$ time.

For \textit{Type-I} sub-query, which is a match of simplex triple pattern, \textit{OP\textsubscript{sp}} is implemented  in a \textit{scan-filter} fashion. The \textit{scan} is a scan of a pattern-specific permutation index. In \textit{filter}, $CR(v_o)$ is derived by a bit-wise AND between the \textit{CandidateRLVertex} result \textit{BitVector} of $v_o$ and the partition's \textit{pVector}. For $CR(v_s)$, we distinguish between two cases. If current \textit{OP\textsubscript{sp}} corresponds to the last matched vertex in the match \textit{order}(discuss in detail in \textit{Section} \ref{subsec:sgdq-algo}), denoted as the \textit{last} \textit{OP\textsubscript{sp}} for brevity, $CR(v_s)$ is derived by a range selection of the \textit{CandidateRLVertex} result \textit{BitVector} of $v_s$. If \textit{OP\textsubscript{sp}} is not the \textit{last} \textit{OP\textsubscript{sp}}, $CR(v_s)$ is derived the same way as $CR(v_o)$. Both  range data and \textit{pVector} is stored as metadata in a partition. By only keep the triples that have \textit{Subject} in the overlapping-free \textit{original partition}, duplications in the final results can be avoided.

For \textit{Type-II} sub-query, we implement \textit{OP\textsubscript{sp}} in a \textit{scan-filter-join-verify} fashion. In essence, as each $\mathcal{P}_j$ is managed by a standalone RDF-3X triple store, we can utilize the query mechanism provided by RDF-3X for \textit{OP\textsubscript{sp}}. More precisely, RDF-3X uses a \textit{scan-join} mechanism for query processing. A query is decomposed into a set of triple patterns. The candidate set of each triple pattern is generated by a scan of a corresponding permutation index. All candidate sets are joined following an optimized order and get the final query results. We further add a \textit{filter} operation before join really happens. At \textit{filter} stage, $CR(v_s)$( and $CR(v_o)$) is derived by bit-wise AND between the \textit{CandidateRLVertex} result \textit{BitVector} of $v_s$(and $v_o$) and the partition's \textit{pVector}. The \textit{filter} introduces \textit{A-Pattern}s to filter irrelevant triple from participate in join operation, this design optimize the  original \textit{scan-join} processing. To avoid final results duplication, we introduce a \textit{verify} stage right after \textit{join} process to prune the results of \textit{scan-filter-join} processing if current \textit{OP\textsubscript{sp}} is the \textit{last} \textit{OP\textsubscript{sp}}. The basic idea behind \textit{verify} is that it keeps the sub-query results that tend to be generated by this \textit{OP\textsubscript{sp}}. This tendency is measured by that the result graph in $\mathcal{G}_R$ have more vertices in the \textit{original partition} generated by $\mathbb{P}$\textsubscript{\textit{METIS}}. It is rational on that $\mathbb{P}$\textsubscript{\textit{METIS}} follows an iterative \textit{coarsen-partition} processing scheme, complete graph $K_n, n\geq 4$ are more likely to be treated as whole in each \textit{coarsen} stage. This means that $K_n, n\geq 4$ are always allocated in a partition. To deal with exceptions, we assume that at most one vertex in a $K_n$ is involved in partition. To illustrate, for $K_3$, which is a triangle, we keep the results of \textit{OP\textsubscript{sp}} that have two vertices in its \textit{original partition}. And for $K_4$, we keep the results that have three vertices in its \textit{original partition}. This is implemented as an range check of all \textit{VertexID} of a result using the range information in the metadata of this partition.

As each \textit{Type-II} sub-query needs to determine its own query plan, the total time of query plan selection can sum up to a high cost. In RDF query processing, cardinality estimation is difficult for the lack of schema. A heuristic-based query plan needs less time to find a sub-optimal plan. We adopt a heuristic-based query plan selection method for each \textit{OP\textsubscript{sp}} follows \cite{heuristics-edbt2012}. Considering the size of data partition, a sub-optimal plan is still acceptable for an \textit{OP\textsubscript{sp}}.

%
%
%
%
\section{Summary Graph Driven Query(SGDQ) Processing} \label{sec:sgdq}

In this section, we explain how SGDQ use summary graph match process to direct \textit{OP\textsubscript{sp}} processing.

\subsection{Summary Graph Match}\label{subsec:sgdq-algo}

According to \textit{definition} \ref{def:match}, the summary graph match is a special case of subgraph isomorphism problem. This problem can be solved using a generic backtracking method\cite{genericbacktracking}. \textit{Algorithm} \ref{alg:subgraphmatch} outlines an implementation of backtracking-based summary graph match. It determines a match order of vertices in $\mathcal{TG}^Q$(\textit{Line} \ref{alg-match:line:order}), then recursively calls a match procedure to find matched $\mathcal{TG}^Q$ in $\mathcal{SG}$. The match process directs the processing of \textit{Op\textsubscript{sp}}(\textit{Line} \ref{alg-match:line:match-start}-\ref{alg-match:line:match-end}), and outputs query results of $Q$ on $\mathcal{G}$. As SGDQ follows the general backtracking paradigm, we only explain the specific features used in SGDQ.

\begin{algorithm}
\footnotesize
\caption{SGDQ processing.}
\label{alg:subgraphmatch}
\KwIn { A query $Q$, $\mathcal{D}$, with $\mathcal{D}_A$ in \textsf{ABIdx}\textsubscript{POS}, \textsf{ABIdx}\textsubscript{comp} and \textsf{ABIdx}\textsubscript{PSO} indexes, and $\mathcal{D}_R$ as $\mathcal{P} = \{\mathcal{P}_1, \ldots, \mathcal{P}_n\}$, summary RDF graph $\mathcal{SG}$ of $\mathcal{D}$. }
\KwOut{ All results of $Q$ on $\mathcal{D}$.}
\nonl \Begin{
\scriptsize{}
Construct $\mathcal{TG}^Q=\langle V^Q_G, E^Q_G\rangle$, $V^Q_G=\{Q^p_1,\ldots,Q^p_m\}$ \; \label{alg-match:line:decomp}
$CR(v)$ =GetCandidateRLVertex($\forall v \in V^Q_G$  $\forall$\textit{A-Patterns}$\in Q$ )\; \label{alg-match:line:crl}
$order$ = \textbf{\textit{DetermineMatchOrder}}($\mathcal{TG}^Q$,$\mathcal{SG}$)\;  \label{alg-match:line:order}
\textbf{\textit{SubgraphMatch}}($\mathcal{TG}^Q$,$\mathcal{SG}$,$M$,$order$,$1$)\;
RetrieveAttribute(); \quad //special steps for \textit{Pattern-II} processing\;
\footnotesize{}
}
\SetKwFunction{SubgraphMatch}{SubgraphMatch}
\SubgraphMatch{$\mathcal{TG}^Q$,$\mathcal{SG}$,$M$,$order$,$d$}{
\scriptsize{}
   \If {$d == |V^Q_T|$ }{                                          \label{alg-match:line:match-start}
        report $R$; \quad //$R$ stores intermediate join results\;  \label{alg-match:line:output}
   }
   $v_s$ = GetNextCandidate($order$)\;                     \label{alg-match:line:nextvertex}
   $C(v_s,M(v_{prev}))$ = \textbf{\textit{ExploreCandidates}}($\mathcal{TG}^Q$,$\mathcal{SG}$,$v_s$,$CR(v_s)$,$order$)\;  \label{alg-match:line:candidate}
   \ForEach {$v \in C(v_s,M(v_{prev}))$ , $v$ is not matched }{   \label{alg-match:line:iter-start}
  		$R_v \leftarrow v_s(v)$ ;\quad //\textit{Op\textsubscript{sp}} of $v_s$ on partition $v$.\;  \label{alg-match:line:opsqmd}
        \If{$R_v = \varnothing$} {
            $CR(v_s) = CR(v_s)- v$; continue \; \label{alg-match:line:opsp-null}
        }
        $R$= \textit{Op\textsubscript{ipj}}$(\mathbb{S}(d-1),R_v)$ \; \label{alg-match:line:opipj}
        \If{$R = \varnothing $} {
            continue \;
        }
		$\mathbb{S} \leftarrow R$; \quad ForwardTrack($\mathbb{S}$,$v_s$,$v$)\; \label{alg-match:line:forward}
		\textbf{\textit{SubgraphMatch}}($\mathcal{TG}^Q$,$\mathcal{SG}$,$M$,$order$,$d+1$)\;  \label{alg-match:line:recursivecall}  		
   		$\mathbb{S} = \mathbb{S} - R$; \quad BackTrack($\mathbb{S}$,$v_s$,$v$)\;   \label{alg-match:line:iter-end}
    }
  \KwRet\;  \label{alg-match:line:match-end}
\footnotesize{}
}
\end{algorithm}

\noindent \textbf{GetCandidateRLVertex}(\textit{Line} \ref{alg-match:line:crl}) process all \textit{A-Pattern}s in $Q$ grouped by common \textit{Subject}. For each query \textit{RL-graph} vertex $v \in V^Q_R$,  \textit{GetCandidateRLVertex} gets the the candidate vertices in $G_R$ as a \textit{BitVector}($v$). If $v$ is a variable with no connected \textit{A-Pattern}s, all bits in \textit{BitVector}($v$) are assigned to 1. If $v$ is an URI of a \textit{VertexID}, the corresponding bit in \textit{BitVector}($v$) are assigned to 1, and 0 elsewhere. Otherwise, a \textit{CandidateRLVertex} operation is called to process all \textit{A-Pattern}s that connect to $v$. \qed

\noindent \textbf{DetermineMatchOrder}(\textit{Line} \ref{alg-match:line:order}). In principle, the \textit{order} is determined as an order of DFS graph traversal in $\mathcal{TG}^Q$. Thus the selection of the starting vertex in  $\mathcal{TG}^Q$ is important. We adopt heuristics that considers the output of \textit{GetCandidateRLVertex} and the the statistics of $\mathcal{D}$. These heuristics are listed as follows, and can be used in combination or separately.

\textit{Heuristic 1}. Choose the order according to the selectivity of all $Q^p_i \in V^Q_T, 1 \leq i$. The selectivity of $Q^p_i$ is roughly estimated as the smallest candidate set size of $v \in Q^p$ that has the smallest candidate set in $V_R$, or the smallest selectivity of \textit{Predicate}s of all \textit{R-Pattern}s in $Q^p_i$. The former can be derived using \textit{GetCandidateRLVertex} function, and the candidate set of $Q^p_i$ in $\mathcal{SG}$ is determined by a range check of all 1-bits in \textit{BitVector}($v$). The later is based on the statistics of $\mathcal{D}$, which are pre-computed as metadata. The $Q^p_i \in V^Q_T$ with smallest selectivity is chosen as the starting vertex. After starting vertex is selected, in the DFS traversal that follows, high selective vertex goes first.

\textit{Heuristic 2}. Given $Q^p_i \in V^Q_T$ that have been arranged in order, if $Q^p_j \in V^Q_T$ is a neighbour of $Q^p_i$ in $\mathcal{TG}^Q$, and merge-join can be applied between them, then arrange $Q^p_j$ as the next in order.

\textit{Heuristic 3}. If \textit\textit{Heuristic 1} and {2} can not apply, select {Type-II} $Q^p_i$ into order with priority. \qed

\noindent \textbf{SubgraphMatch}(\textit{Line} \ref{alg-match:line:match-start}-\ref{alg-match:line:match-end}) is a recursive function. It starts as gets a current query vertex $v$ from \textit{order}(\textit{Line} \ref{alg-match:line:nextvertex}), and obtains a set of candidate vertices of $M(v)$ in $\mathcal{SG}$, denoted as $C(v,M(v_{prev}))$, which is the set of neighbour vertices\footnote{Considering the existence loop edges for each vertex $\mathcal{SG}$, the neighbours contains the vertex itself.} of the previous matched vertices $M(v_{prev})$. The vertex match in $\mathcal{SG}$ is judged by vertex label match, and corresponding bit of $M(v) \in C(v,M(v_{prev}))$  in \textit{BitVector}($M(v)$) is 1-bit(\textit{Line} \ref{alg-match:line:candidate}).

\textit{SubgraphMatch} iterates over all candidate vertices $M(v) \in C(v,M(v_{prev}))$ (\textit{Line} \ref{alg-match:line:iter-start}-\ref{alg-match:line:iter-end}). In each iteration, \textit{SubgraphMatch} calls an \textit{Op\textsubscript{sp}}, i.e. $v(M(v))$, and gets its results $R(v(M(v)))$(\textit{Line} \ref{alg-match:line:opsqmd}). If $R(v(M(v))) = \varnothing$, $v$ is removed by setting the corresponding bit in \textit{BitVector}($v$) to 0(\textit{Line} \ref{alg-match:line:opsp-null}). This avoids unnecessary repeated execution of \textit{Op\textsubscript{sp}} that do not satisfy vertex math in \textit{Definition} \ref{def:match}. For each edge between $v$ and all $v_{prev}$ in $\mathcal{TG}^Q$, an \textit{Op\textsubscript{ipj}}$(v(M(v),v_{prev}(M(v_{prev})))$ is called to check for edge injection(\textit{Line} \ref{alg-match:line:opipj}). As the search order follows a $\mathcal{TG}^Q$ graph traversal order, in most cases \textit{Op\textsubscript{ipj}} are implemented using hash join. We use order-preserving merge-joins for \textit{Op\textsubscript{ipj}} if possible, considering \textit{Heuristic 2} in \textit{DetermineMatchOrder}. If the result set $R(v(M(v),v_{prev}(M(v_{prev})))$ is not $\varnothing$, the join results are cached in a structure $\mathbb{S}$(\textit{Line} \ref{alg-match:line:forward}). $\mathbb{S}$ enables backtracking to previous matched vertex, and the cached intermediate results in $\mathbb{S}$ can be reused for all vertices in the candidate region $C(v,M(v_{prev}))$.

Next, \textit{SubgraphMatch} recursively calls another \textit{SubgraphMatch} with an incremental in depth $d$(\textit{Line} \ref{alg-match:line:recursivecall}). If $d$ equals to the number of vertex in $\mathcal{TG}^Q$, this means a match of $\mathcal{TG}^Q$ is generated, and currently cached $R$ can be output as part of final results(\textit{Line} \ref{alg-match:line:output}). If \textit{Op\textsubscript{ipj}} returns $\varnothing$, \textit{SubgraphMatch} iterate to the next candidate if exists, otherwise backtrack to the next candidate vertex in $\mathcal{SG}$ for $v_{prev} \in \mathcal{TG}^Q$ in order. \qed

\begin{property}[\textbf{Query Completeness and Non-Redundancy}]
Let $R_{SGDQ}$ and $R$ be the result set of  generated by $Q$ on $\mathcal{D}$ and  SGDQ respectively. If $r \in R$, then $r \in R_{SGDQ}$ holds.
\end{property}

This property can be proofed by contradiction. We omit the details in this paper.
%
%
%

\subsection{Cost Analysis}

Given $\mathcal{D}_A$ , and $\mathcal{D}_R$ as $\mathcal{P}$ of $n$ partitions. For a query $Q$, assume that in $Q$ there are totally $t_{I}$ \textit{Pattern-I A-Pattern}s, $t_{II}$ \textit{Pattern-II A-Pattern}s, and $Q$ is decomposed into $m$ sub-queries, of which $m_I$ \textit{Type-I}, and $m_{II}$ \textit{type-II}. The average cost of \textit{Type-I} and \textit{Type-II} sub-query are $Cost(\textit{Op\textsubscript{sp}})$ and $Cost(\textit{Op\textsubscript{sp}})$ respectively.

We first give the cost analysis of \textit{A-Patterns}. As there are total $|\mathcal{D}_A|$ \textit{BitVector}s, the cost of \textit{CandidateRLVertex} operation is the time complexity of index lookup using a composite key of \textit{p} and \textit{o}, which is $O(log|\mathcal{D}_A|)$. Similarly, the cost of \textit{RetrieveAttribute} operation is $O(log|V_R|)$. Consequently the aggregated cost of all \textit{A-Pattern}s processing sum up to $O(t_I\cdot (log|\mathcal{D}_A|)+ t_{II}\cdot log|V_R|)$。

The \textit{R-Pattern}s processing follows summary graph matching paradigm. Conventional optimization methods of subgraph matching are focused on establishing graph topological indexes to effectively prune candidate vertices. Such optimizations can not work for \textit{Algorithm} \ref{alg:subgraphmatch}, as we still needs to actually take joins to determine whether two partitions are joinable. Thus exhaustive exploration of all feasible neighbours is adopted in \textit{Algorithm} \ref{alg:subgraphmatch}. This is acceptable as $\mathcal{SG}$ size is relative small. To this end, In worst case SGDQ needs $\sum^{m-1}_{i=1}|V_S|^{i}$ times of \textit{Op\textsubscript{sp}}, and $|E_S|\cdot\sum^{m-1}_{i=1}|V_S|^{i}$ times of \textit{Op\textsubscript{ipj}}. The cost of \textit{R-Pattern}s processing can sum up to $(\sum^{m-1}_{i=1}|V_S|^{i})\cdot Cost(\textit{Op\textsubscript{sp}}) + (|E_S|\cdot\sum^{m-1}_{i=1}|V_S|^{i}) \cdot Cost(\textit{Op\textsubscript{ipj}})$.

SGDQ only needs to maintain intermediate result at each recursive call, which is generated on $m$ partitions. This significantly reduces the cost to keep vast amount of immediate results generated by scans and joins on large dataset.

%
%
%
%

\section{Related Works}\label{sec-related}

RDF query optimization has attracted interests from both industrial and research domain, and considerable methods, either relational or graph-based, centralized or distributed, have been proposed. In this section, we only list the most related to our approach and compare them with ours.

\noindent \textbf{Graph-based approaches.} RDF query in native graph storage is implemented as subgraph matching. Graph pattern matching has been extensively studied\cite{genericbacktracking,gpm-02}. Following these, \cite{trinitrdf-vldb2013,subgraph-vldb2015} use fine-grained graph exploration for RDF query. This need to maintain the entire graph in memory. \cite{gstore-vldb2011,structuralindex-tkde2013} introduce structural summaries of graph as disk-based structural index. Matched results are generated using a filter-and-refinement strategy. SGDQ constructs a summary graph based on graph partitions, which is most similar to \cite{triad-sigmod2014}. SGDQ differs from \cite{triad-sigmod2014} in that the summary graph is used in SGDQ for arrange sub-query processings, while it is used in \cite{triad-sigmod2014} for pruning unnecessary data before join processing.

\noindent \textbf{Relational approaches.} A na\"{\i}ve way to manage RDF data is adopting relational database to store all triples in a single 3 column table. Query is implemented in a \textit{scan-join} fashion\cite{rdf3x2010,hexastore-vldb2008,triad-sigmod2014}. Optimizations of na\"{\i}ve approach are mainly from three aspects. \textbf{i)}\textbf{Using permutation index to make join and scan themselves efficient.} \cite{permuindex-laweb2005} first introduce the concept of permutation index. Hexastore\cite{hexastore-vldb2008} exhaustively indexes all SPO permutations and stores them directly in sorted B+-tree. RDF-3X\cite{rdf3x2010} further introduces extra indexes of all SPO attributes projections. Such permutation indexes enables efficient range scan and merge-join at the cost of data redundancy. SGDQ extensively use permutation indexes in manage each partitions. \textbf{ii)}\textbf{Reducing the number of joins needed.} Property table\cite{jena} clustering relevant \textit{predicate}s together in a table schema. This helps in eliminating \textit{s-s join}s, but requires a previous knowledge about query. BitMat\cite{bitmat-www2010} represents property tables as 3 dimension bitmaps, and replace joins with bitwise operations which share common elements at a dimension. SGDQ is different in that it classifies triples and adopts bitmaps only for \textit{A-Triple}s. This eliminates all \textit{s-s join}s related to \textit{A-Patten}s, and avoids complex considerations of other types of joins when using bitmap index. \textbf{iii)}\textbf{Filtering out irrelevant inputs of join}. RDF-3X adopted a technique named SIP(Sideway Information Passing) to dynamically determine the unnecessary data blocks that can be skipped during scan\cite{rdf3x2009}. RG-index\cite{rg-index} introduces a relational operator \textit{filter} that using frequent graph pattern to prune triples that can not match. SGDQ uses filter in each \textit{Op\textsubscript{sp}} and shares the same objective as \cite{rdf3x2009,rg-index}, but it uses the results of \textit{A-Pattern}s processing in filtering.

\noindent \textbf{Data partition} are widely used in shared-nothing distribute RDF systems\cite{1UHC} \cite{triad-sigmod2014}. They follow the principles of evenly distribute data to each computing node so that each partitions can be processed in parallel to boost the overall query performance. The number of partitions are generally equal to the number of computing nodes, and the goal of optimization is to design effective partition strategy to minimize the cost of distribute inter-partition joins. To this end, the most common adopted partition strategies are edge-based hash\cite{hashpartition} and graph partition\cite{1UHC}. SGDQ adopts partition in a totally different way. It partition data in order to decompose original query into a series of light-weight processing, the number of partitions depends on the dataset size.

%
%
%
%

\section{Evaluation}\label{sec-evaluation}

In this section, we empirically evaluate the RDF query processing performance on two kinds of large-scale benchmark data. The goals of our evaluations are \textbf{i)} We show the correctness and the competitive performance of SGDQ on basic benchmark queries (contain at most 6 triple patterns) over the state-of-the-art \textit{triple store}s(\textit{Section} \ref{subsec:eval:basic}); \textbf{ii)} We show the superior performance of SGDQ on complex queries that involve up to 16 triple patterns, and analyze the effect of our methods(\textit{Section} \ref{subsec:eval:lubm-complex}, \ref{subsec:eval:snib-complex}).

\subsection{Environment, Competitors and Datasets}

All experiments were performed on a server with Debian 7.4 in 64-bit Linux kernel, two Intel Xeon E5-2640 2.0GHz processors and 64GB RAM. The query response time was measured in milliseconds. SGDQ is developed in C++, and is compiled using GCC-4.7.1 with -O3 optimization flag. We used METIS 5.1\footnote{\url{http://glaros.dtc.umn.edu/gkhome/fetch/sw/metis/metis-5.1.0.tar.gz}} to get the \textit{original partition} of RDF graph, and RDF-3X 0.3.8\footnote{\url{https://rdf3x.googlecode.com/files/rdf3x-0.3.8.zip}} for partition management and support of sub-query processing. For fair comparison, We also used the dictionary to convert between the string value of \textit{s},\textit{o},\textit{p} and the identifiers used in processing, and presented results in its original string value form.

\noindent \textbf{Competitors}. There exists a wide choice on the state-of-the-art \textit{triple store}. We chosen three publicly available and representative ones. We list these competitors as:

\textbf{RDF-3X} maintains all possible permutation of \textit{s},\textit{p} and \textit{o} as indexes, and use B+-tree to facilitate index lookup. Query of a single pattern is implemented as an index scan operation, and merge join are extensively leveraged to join between two patterns. Together with index-specific query optimization techniques\cite{rdf3x2009,rdf3x2010}, RDF-3X remains a competitive state-of-the-art \textit{triple stores}.

\textbf{TripleBit}\footnote{\url{http://grid.hust.edu.cn/triplebit/TripleBit.tar.gz}} organizes data in a compact bit matrix, and vertically partition the matrix into sub-matrices following predicates dimension and sorted subject/object dimension. Such compact data representation and partition can minimizes the size of intermediate results, thus theoretically results in a query performance promotion\cite{triplebit-vldb2013}.

\textbf{Virtuoso 7}\footnote{\url{https://github.com/openlink/virtuoso-opensource}} is a commercial-of-the-shelf column store that support RDF management. It exploit existing relational RDBMS techniques and add functionality to deal with features specific to RDF data. It also adopted bitmap index to take advantage of the property that many triple share the same predicate and object\cite{virtuoso-bitmap}.

\noindent \textbf{Dataset}. We adopted two different kinds of benchmarks.

\textbf{LUBM}\footnote{\url{http://swat.cse.lehigh.edu/projects/lubm/}} is a widely used benchmark in both academical researches and industrial applications. It provides a university database where the components, such as university, departments, professors, students etc, can be polynomially generated. The size of dataset scales according to the number of universities. We generated two dataset, they are identified as LUBM-8000 and LUBM-20480 according to the number of universities. As all components in the data are generated in a proportional fashion, the RDF graph of a LUBM dataset can be regarded as a random graph.

\textbf{SNIB}\footnote{\url{http://www.w3.org/wiki/Social_Network_Intelligence_BenchMark}} provides the semantic dataset of a twitter-like social network, which includes resources like users, post, reply, tags and comments, etc. The dataset scales according to the number of users. We generate a SNIB dataset of 15000 users using S3G2\cite{s3g2-generator}, identified as SNIB-15000. As it models a social network, the RDF graph of SNIB-15000 is a power-law graph.

Remind that to support the original LUBM queries, the standard way is to inference on RDF Schema \textit{subclasses} and \textit{subproperties}. We used the inference engine in Virtuoso 7 in our experiment, and the inferred triples were also managed by each triple store. The statistics of dataset are listed in \textit{Table} \ref{tbl:datastats}.

\begin{table}[ht]
\captionsetup{belowskip=0pt,aboveskip=0pt}
\caption{Statistics of data collection .}
\label{tbl:datastats}
\scriptsize
\centering   
\begin{tabular}{ | >{\centering\arraybackslash}m{0.56in} | >{\centering\arraybackslash}m{0.48in} |  >{\centering\arraybackslash}m{0.48in} | >{\centering\arraybackslash}m{0.42in} | >{\centering\arraybackslash}m{0.48in} |>{\centering\arraybackslash}m{0.08in} |}
    \hline
    \multirow{2}{*}{Dataset} & \multicolumn{2}{ c |}{\# Triples, $|\mathcal{D}|$ } & \multicolumn{2}{ c |}{RL-Graph, $\mathcal{G}_R$} & \multirow{2}{*}{\#P}  \\
    \cline{2-5}
    \multirow{2}{*}{} & \# Totals   & \# Inferred & $|V|$ & $|E|$ &  \multirow{2}{*}{} \\
    \hline
    LUBM-8000   &   1,970,947,330   &   869,030,729   &   173,806,051   &   529,614,087  & 18   \\
    \hline
    LUBM-20480   &   4,844,909,074   &   2,024,718,666   &   444,851,202   &   1,355,446,182  & 18   \\
    \hline
    SNIB-15000   &   1,503,766,205   &   27,618,900   &  125,779,052   &   875,721,431  & 48   \\
    \hline
\end{tabular}
\end{table}

\noindent \textbf{Queries}. For LUBM dataset, we used the 14 queries provided by LUBM benchmark. Furthermore, we manually constructs 6 complex queries on LUBM and and 7 queries on SNIB dataset. These complex queries contains more triple patterns than the original ones. All Queries are listed in the \textit{Appendix}\label{sec-appendix}.

Consider that the cache mechanism of operating system have a profound impact in query execution, we measure the performance using both cold cache mode and warm cache mode. In cold cache mode, we purged the OS file system cache each time before running a query\footnote{In Debian OS, we purged its file system cache using commands: \textbf{/bin/sync; echo 3 $>$ /proc/sys/vm/drop\_caches}}. Each query was executed in cold cache mode and warm cache mode for 10 times in a successive manner. The result in each mode was reported as the arithmetic mean and was round to millisecond\footnote{Actually, in warm cache mode, a query was executed 11 times consecutively. The first execution was for cache warm-up, and its result was removed from averaging. }.

\begin{table*}[ht]
\captionsetup{belowskip=0pt,aboveskip=0pt}
\caption{Benchmark query response time on LUBM dataset(in milliseconds).}
\captionsetup[sub]{margin=0pt,skip=0pt, labelfont=bf, labelformat=parens, textfont=normalfont}
\scriptsize
\label{tbl:lubm}
\centering

  $^{\bigstar}$: Wrong in query result size.
  \quad \quad \textbf{N/A}: Query does not end in 12 hours.

\vspace{\abovedisplayskip}

\begin{minipage}[b]{\textwidth}
\subcaption{LUBM-8000 dataset.}
\begin{tabular}{ | >{\centering\arraybackslash}m{0.40in}  ? >{\raggedleft\arraybackslash}m{0.3in} | >{\raggedleft\arraybackslash}m{0.3in} |>{\raggedleft\arraybackslash}m{0.3in} |>{\raggedleft\arraybackslash}m{0.3in} |>{\raggedleft\arraybackslash}m{0.3in} |>{\raggedleft\arraybackslash}m{0.3in} |>{\raggedleft\arraybackslash}m{0.3in} |>{\raggedleft\arraybackslash}m{0.3in} |>{\raggedleft\arraybackslash}m{0.3in} |>{\raggedleft\arraybackslash}m{0.3in} |>{\raggedleft\arraybackslash}m{0.3in} |>{\raggedleft\arraybackslash}m{0.3in} |>{\raggedleft\arraybackslash}m{0.3in} |>{\raggedleft\arraybackslash}m{0.3in} | }
\hline
        & \multicolumn{1}{c|}{L1} & \multicolumn{1}{c|}{L2}  & \multicolumn{1}{c|}{L3} & \multicolumn{1}{c|}{L4} & \multicolumn{1}{c|}{L5} & \multicolumn{1}{c|}{L6} & \multicolumn{1}{c|}{L7} & \multicolumn{1}{c|}{L8} & \multicolumn{1}{c|}{L9} & \multicolumn{1}{c|}{L10} & \multicolumn{1}{c|}{L11} & \multicolumn{1}{c|}{L12} & \multicolumn{1}{c|}{L13} & \multicolumn{1}{c|}{L14}  \\[0.2ex]
\hline
\multicolumn{15}{|c|}{\textbf{Cold Cache.}} \\
\hline    
RDF-3X      &   377     &   1,044,390 &   372   &   456   &   377   &   2,508,477   &   541   &   1,040  &   517,775    &   385   &   497   &   440   &   24,479   &   1,828,117   \\
\hline    
Virtuoso7   &  {\cellcolor[gray]{0.8}\textbf{7}}  &   25,193   &   3    &  {\cellcolor[gray]{0.8}\textbf{7}}   &   {\cellcolor[gray]{0.8}\textbf{5}}   &   2,450,552 &   {\cellcolor[gray]{0.8}\textbf{4}}   &   628    &   225,441  &   2   &   5   &   55 & 4,381   &   1,981,602  \\
\hline    
TripleBit   &  115   &   91,085$^{\bigstar}$   &   {\cellcolor[gray]{0.8}\textbf{2}}    &  62   &   51   &   3,891,150 &   37   &   109$^{\bigstar}$    &   423,428  &   {\cellcolor[gray]{0.8}\textbf{0.3}}   &   {\cellcolor[gray]{0.8}\textbf{3}}   &   {\cellcolor[gray]{0.8}\textbf{5}} & 626   &   3,122,737 \\
\hline    
SGDQ        &  92   &   {\cellcolor[gray]{0.8}\textbf{11,790}}   &   42    &  56   &   49   &   {\cellcolor[gray]{0.8}\textbf{924,535}} &   102   &   {\cellcolor[gray]{0.8}\textbf{325}}    &   {\cellcolor[gray]{0.8}}\textbf{94,702}  &   22   &   63  &   92 & {\cellcolor[gray]{0.8}\textbf{271}}  &   {\cellcolor[gray]{0.8}\textbf{592,374}} \\
\hline
\multicolumn{15}{|c|}{\textbf{Warm Cache.}} \\
\hline    
RDF-3X  &  2   &   910,820   &   2    &  5   &   7   &   2,196,851 &   7   &   439    &   109,162  &   2   &   186   &   19 & 6,174   &   1,553,394 \\
\hline    
Virtuoso7   &  2   &   24,033   &   1    &  3   &   4   &   2,381,088 &   2   &   427    &   223,551  &   1   &   2   &   21 & 1,592   &   1,568,623 \\
\hline    
TripleBit   &  {\cellcolor[gray]{0.8}\textbf{0.2}}   &   39,337$^{\bigstar}$   &   {\cellcolor[gray]{0.8}\textbf{0.3}}    &  3   &   3   &   3,823,121 &   2   &   102$^{\bigstar}$    &   178,381  &   {\cellcolor[gray]{0.8}\textbf{0.2}}   &   2   &   {\cellcolor[gray]{0.8}\textbf{4}} & 582   &   3,045,244 \\
\hline    
SGDQ    &  2   &   {\cellcolor[gray]{0.8}\textbf{8,015}}   &   1    &  {\cellcolor[gray]{0.8}\textbf{2}}   &   {\cellcolor[gray]{0.8}\textbf{3}}   &   {\cellcolor[gray]{0.8}\textbf{892,722}} &   {\cellcolor[gray]{0.8}\textbf{2}}   &   {\cellcolor[gray]{0.8}\textbf{221}}    &   {\cellcolor[gray]{0.8}\textbf{76,118}}  &   1   &   {\cellcolor[gray]{0.8}\textbf{2}}   &   32 & {\cellcolor[gray]{0.8}\textbf{64}}   &   {\cellcolor[gray]{0.8}\textbf{573,189}} \\
\hline    
\hline    
\#Results    &  4   &   2528   &   6    &  34   &   719   &   \tiny{83,557,706} &   67   &   7,790    &   2,178,420  &   4   &   224   &   15 & 37,144   &   \tiny{63,400,587} \\
\hline    
\end{tabular}
\label{subtbl:lubm8000}
\end{minipage}

\vspace{\belowdisplayskip}

\begin{minipage}[b]{\textwidth}
\subcaption{LUBM-20480 dataset.}
\begin{tabular}{ | >{\centering\arraybackslash}m{0.40in}  ? >{\raggedleft\arraybackslash}m{0.3in} | >{\raggedleft\arraybackslash}m{0.3in} |>{\raggedleft\arraybackslash}m{0.3in} |>{\raggedleft\arraybackslash}m{0.3in} |>{\raggedleft\arraybackslash}m{0.3in} |>{\raggedleft\arraybackslash}m{0.3in} |>{\raggedleft\arraybackslash}m{0.3in} |>{\raggedleft\arraybackslash}m{0.3in} |>{\raggedleft\arraybackslash}m{0.3in} |>{\raggedleft\arraybackslash}m{0.3in} |>{\raggedleft\arraybackslash}m{0.3in} |>{\raggedleft\arraybackslash}m{0.3in} |>{\raggedleft\arraybackslash}m{0.3in} |>{\raggedleft\arraybackslash}m{0.3in} | }
\hline
        & \multicolumn{1}{c|}{L1} & \multicolumn{1}{c|}{L2}  & \multicolumn{1}{c|}{L3} & \multicolumn{1}{c|}{L4} & \multicolumn{1}{c|}{L5} & \multicolumn{1}{c|}{L6} & \multicolumn{1}{c|}{L7} & \multicolumn{1}{c|}{L8} & \multicolumn{1}{c|}{L9} & \multicolumn{1}{c|}{L10} & \multicolumn{1}{c|}{L11} & \multicolumn{1}{c|}{L12} & \multicolumn{1}{c|}{L13} & \multicolumn{1}{c|}{L14}  \\[0.2ex]
\hline
\multicolumn{15}{|c|}{\textbf{Cold Cache.}} \\
\hline    
RDF-3X  &   1,327 &   2,383,700 &   377   &   533   &   394   &   6,265,536   &   511   &   1,028  &   843,811    &   387   &   1,845   &   497   &   72,766   &   4,310,605   \\
\hline    
Virtuoso7    &  {\cellcolor[gray]{0.8}\textbf{7}}   &   41,169   &   {\cellcolor[gray]{0.8}\textbf{5}}    &  {\cellcolor[gray]{0.8}\textbf{8}}   &   {\cellcolor[gray]{0.8}\textbf{5}}   &   6,661,334 &   {\cellcolor[gray]{0.8}\textbf{59}}   &   888    &   425,757  &   9   &   8   &   19 & 4,291   &   5,587,519 \\
\hline    
TripleBit  &  124   &   187,712$^{\bigstar}$   &   98    &  112   &   89  &   8,224,312 &   137   &   182$^{\bigstar}$    &   739,002  &   {\cellcolor[gray]{0.8}\textbf{1}}   &   {\cellcolor[gray]{0.8}\textbf{3}}   &   {\cellcolor[gray]{0.8}\textbf{6}} & 1,127   &   7,811,369 \\
\hline    
SGDQ    &  76   &   {\cellcolor[gray]{0.8}\textbf{24,327}}   &   64    &  75   &   68  &   {\cellcolor[gray]{0.8}\textbf{1,224,376}} &   112   &   {\cellcolor[gray]{0.8}\textbf{122}} &  {\cellcolor[gray]{0.8}\textbf{172,411}} &   21   &   72   &  107 & {\cellcolor[gray]{0.8}\textbf{433}}   &   {\cellcolor[gray]{0.8}\textbf{1,502,331}} \\
\hline
\multicolumn{15}{|c|}{\textbf{Warm Cache.}} \\
\hline    
RDF-3X  &   2   &   2,330,893 &   3   &   5   &   7   &   5,781,561 &   7   &   401 &   830,944  &   2   &   362  &  14   &   38,988   &   3,813,701 \\
\hline    
Virtuoso7  &  2   &   42,521   &   1    &  {\cellcolor[gray]{0.8}\textbf{3}}   &   4   &   6,253,872 &   4  &   402    &   440,644  &   3   &   3   &   24 & 3,759   &   4,532,172 \\
\hline    
TripleBit  &  {\cellcolor[gray]{0.8}\textbf{0.2}}   &   56,005$^{\bigstar}$   &   {\cellcolor[gray]{0.8}\textbf{0.3}}    &  2   &   {\cellcolor[gray]{0.8}\textbf{3}}   &   8,102,716 &   5   &   169$^{\bigstar}$    &   527,032  &   {\cellcolor[gray]{0.8}\textbf{0.5}}   &   3   &   {\cellcolor[gray]{0.8}\textbf{4}}   &   639   &   7,692,721 \\
\hline    
SGDQ    &  2   &   {\cellcolor[gray]{0.8}\textbf{12,073}}   &   2    &  5   &   4   &   {\cellcolor[gray]{0.8}\textbf{1,123,549}} &   {\cellcolor[gray]{0.8}\textbf{4}}   &   {\cellcolor[gray]{0.8}\textbf{23}}    &   {\cellcolor[gray]{0.8}\textbf{122,395}}  &   1   &   {\cellcolor[gray]{0.8}\textbf{2}}   &  17 & {\cellcolor[gray]{0.8}\textbf{117}}   &   {\cellcolor[gray]{0.8}\textbf{1,422,801}} \\
\hline    
\hline    
\#Results    &  4   &   2,528   &   6    &  34   &   719   &   \tiny{213,817,916} &   67   &   7,790    &   2,703,043  &   4   &   224   &   15 & 95,522   &   \tiny{162,211,567} \\
\hline    
\end{tabular}
\label{subtbl:lubm20480}
\end{minipage}

\end{table*}

\subsection{Effects of Partitions}

We adopted $n=500$, i.e. the RL-graph of each dataset was partitioned into 500 subgraphs using METIS. The effects of $\mathbb{P}$\textsubscript{\textit{METIS}} and $\mathbb{P}$\textsubscript{\textit{1-UHC}} is given in \textit{Table} \ref{tbl:partition}, manifested as the number of cutting edges and \textit{replication factor} $\alpha$ respectively. By managing \textit{A-Triples} individually, $\mathbb{P}$\textsubscript{\textit{1-UHC}} have far smaller duplicated triples compared with that of the \textit{un-two} strategy show in \cite{1UHC}.

\begin{table}[ht]
\captionsetup{belowskip=0pt,aboveskip=0pt}
\caption{Partition characteristics }
\label{tbl:partition}
\scriptsize
\centering   
\begin{tabular}{ | >{\centering\arraybackslash}m{0.56in} | >{\centering\arraybackslash}m{1.2in} |  >{\centering\arraybackslash}m{0.4in} |}
    \hline
    Dataset &   \# cutting edges   &   $\alpha$ \\
    \hline
    LUBM-8000   &   23,624,351   &   1.23      \\
    \hline
    LUBM-20480   &   61,518,672   &   1.21     \\
    \hline
    SNIB-15000   &   58,823,356   &   1.52      \\
    \hline
\end{tabular}
\end{table}

We list the physical storage characteristics in \textit{Table} \ref{tbl:storagestats}\footnote{Virtuoso 7 manage all dataset in a single database instance. The storage characteristic of Virtuoso 7 is measured by the increased database file size after each dataset is loaded.}. Basically, RDF-3X costs the most space, as it needs to maintain all 6 SPO permutation indexes and 9 aggregated indexes of binary and unary projection. Even though SGDQ use RDF-3X to manage all partitions of \textit{R-Triples}, but the bitmap index of \textit{A-Triple}s contributes to a more compact storage. Consider the proportion of \textit{A-Triples} in a dataset(as shown in \textit{Table} \ref{tbl:datastats}), SGDQ shows a more efficient storage compared with RDF-3X.

\begin{table}[ht]
\captionsetup{belowskip=0pt,aboveskip=0pt}
\caption{Physical storage characteristics (in GB). }
\label{tbl:storagestats}
\scriptsize
\centering   
\begin{tabular}{ | >{\centering\arraybackslash}m{0.56in} | >{\centering\arraybackslash}m{0.5in} |  >{\centering\arraybackslash}m{0.35in} | >{\centering\arraybackslash}m{0.35in} | >{\centering\arraybackslash}m{0.35in} |>{\centering\arraybackslash}m{0.35in} |}
    \hline
    Dataset &   Raw N3 file   &   RDF-3X  &   Virtuoso7   & TripleBit &   SGDQ  \\
    \hline
    LUBM-8000   &   343   &   124   &   58   &   57  & 92   \\
    \hline
    LUBM-20480   &   848   &   283   &   129   &   147  & 184   \\
    \hline
    SNIB-15000   &   203   &   97   &   42  &   49  & 84   \\
    \hline
\end{tabular}
\end{table}

\subsection{LUBM benchmark queries} \label{subsec:eval:basic}

We report the response time of query on LUBM-8000 and LUBM-20480 dataset in \textit{Table} \ref{subtbl:lubm8000} and \textit{Table} \ref{subtbl:lubm20480} respectively. Generally speaking, SGDQ achieve a comparable query performance in comparison to its competitors using standard benchmark queries.

Query \textit{L1}, \textit{L3}, \textit{L4}, \textit{L5}, \textit{L7}, \textit{L8}, \textit{L10}, \textit{L11} exhibit the same characteristics. They all have a high-selective \textit{A-Pattern}, and their result sets are quite small and are independent of the dataset size. Among them, \textit{L4} is a bit more complex, it contains 3 \textit{Pattern-II} \textit{A-Pattern}s. SGDQ needs an extra bitmap lookup to retrieve the variable of these \textit{A-Pattern}.
Due to their simple nature, they can not benefit from SGDQ optimizations. In fact, the response time of all these queries for both SGDQ and all competitors are within 1 second for both cold cache and warm cache. It is hard for a user to feel the difference. Although Query \textit{L13} also have a high-selective \textit{A-Pattern}, the result set of \textit{L13} increase as the dataset size increase. For cold cache query, RDF-3X needs to scan the large indexes and cache the data at each query time. SGDQ outperforms RDF-3X by 4.1(\textit{L1} on LUBM-8000) to 168(\textit{L13} on LUBM-20480) times. This proves the effectiveness of SGDQ in avoiding the cost of scan large indexes by manage and query data in partitions. As for warm cache query, the candidate triples of high selective patterns can be held in cache. As SGDQ introduce extra cost of bitmap search and partition selection, SGDQ has no advantages over competitors.

For the time consuming queries, i.e. \textit{L2}, \textit{L6}, \textit{L9} and \textit{L14}, SGDQ outperforms all competitors by 1.7(\textit{L2} for Virtuoso 7 on LUBM-20480) to 98(\textit{L2} for RDF-3X on LUBM-20480) times for cold cache, and by 2.3(\textit{L9} for TripleBit on LUBM-8000) to 193(\textit{L2} for RDF-3X on LUBM-20480) times for warm cache. For \textit{L6} and \textit{L14}, SGDQ is efficient in that they only need to look up the bitmap indexes of \textit{A-Triples}, the cost is due to the dictionary operation for string and ID conversation.
For \textit{L2} and \textit{L9},  theirs \textit{RL-graph} is a triangle. This means that they can be answered without inter-partition processing, and theirs $\mathcal{TG}^Q$ have only one vertex. Intuitively, the sum of all \textit{OP\textsubscript{sp}} are far better than execute the query as a whole.

\begin{table*}[ht]
\captionsetup{belowskip=0pt,aboveskip=0pt}
\caption{Complex queries response time on LUBM-20480 dataset(in milliseconds).}
\centering

  $^{\bigstar}$: Wrong in query result size.
  \quad \quad \textbf{N/A}: Query does not end in 12 hours.

\vspace{\abovedisplayskip}
\label{tbl:complex-lubm}
\scriptsize
\begin{tabular}{ | >{\centering\arraybackslash}m{0.40in}  ? >{\raggedleft\arraybackslash}m{0.4in} | >{\raggedleft\arraybackslash}m{0.4in} |>{\raggedleft\arraybackslash}m{0.4in} |>{\raggedleft\arraybackslash}m{0.3in} |>{\raggedleft\arraybackslash}m{0.3in} |>{\raggedleft\arraybackslash}m{0.3in} ? >{\raggedleft\arraybackslash}m{0.4in} |>{\raggedleft\arraybackslash}m{0.4in} |>{\raggedleft\arraybackslash}m{0.4in} |>{\raggedleft\arraybackslash}m{0.3in} |>{\raggedleft\arraybackslash}m{0.3in} |>{\raggedleft\arraybackslash}m{0.3in} | }
    \hline
    \multirow{2}{*}{} & \multicolumn{6}{ c ?}{LUBM-8000} & \multicolumn{6}{ c |}{LUBM-20480}  \\
    \cline{2-13}
    \multirow{2}{*}{} &  \multicolumn{1}{c|}{L15} & \multicolumn{1}{c|}{L16}  & \multicolumn{1}{c|}{L17} & \multicolumn{1}{c|}{L15\textsubscript{sel}} & \multicolumn{1}{c|}{L16\textsubscript{sel}} & \multicolumn{1}{c?}{L17\textsubscript{sel}}  & \multicolumn{1}{c|}{L15} & \multicolumn{1}{c|}{L16}  & \multicolumn{1}{c|}{L17} & \multicolumn{1}{c|}{L15\textsubscript{sel}} & \multicolumn{1}{c|}{L16\textsubscript{sel}} & \multicolumn{1}{c|}{L17\textsubscript{sel}}   \\
\hline
\multicolumn{13}{|c|}{\textbf{Cold Cache.}} \\
\hline    
RDF-3X  & 666,179 & 989,280  & N/A & 3,783 & 14,968 & 83,241 & 1,642,307  & N/A & N/A & 30,810 & 75,919 & 113,143    \\
\hline    
Virtuoso7    & 528,767 & 25,533  & {\cellcolor[gray]{0.8}\textbf{12,492}} & {\cellcolor[gray]{0.8}\textbf{81}} & 731 & 474 & 1,364,557  & 41300 & {\cellcolor[gray]{0.8}\textbf{24,285}} & {\cellcolor[gray]{0.8}\textbf{82}} & 954 & 544     \\
\hline    
TripleBit   & 1,942,030$^{\bigstar}$ & 189,630$^{\bigstar}$  & 95,959$^{\bigstar}$ & 3,815$^{\bigstar}$ & 13,379$^{\bigstar}$ & 25,145 &  1,188,727$^{\bigstar}$    &   535,515$^{\bigstar}$     &  1,643,845$^{\bigstar}$    &   5,900$^{\bigstar}$ &   19,433$^{\bigstar}$   &   9,731    \\
\hline    
SGDQ    & {\cellcolor[gray]{0.8}\textbf{249,532}} & {\cellcolor[gray]{0.8}\textbf{21,727}}  & 16,392 & 122 & {\cellcolor[gray]{0.8}\textbf{322}} & {\cellcolor[gray]{0.8}\textbf{215}} & {\cellcolor[gray]{0.8}\textbf{852,380}} & {\cellcolor[gray]{0.8}\textbf{39,271}}  & 28,183 & 158 & {\cellcolor[gray]{0.8}\textbf{334}} & {\cellcolor[gray]{0.8}\textbf{292}}    \\
\hline
\multicolumn{13}{|c|}{\textbf{Warm Cache.}} \\
\hline    
RDF-3X    & 483,763 & 966,009  & N/A & 1,499 & 6,519  & 47,877 &  1,184,803 & N/A & N/A & 3,783 & 14,968 & 92,775  \\
\hline    
Virtuoso7  & 522,486 & 24,428  & 12,072 & 63 & 383 & 371 & 1,345,170  & 41,013 & 23,626 & {\cellcolor[gray]{0.8}\textbf{83}} & 951 & 444    \\
\hline    
TripleBit  & 1,800,790$^{\bigstar}$ & 171,661$^{\bigstar}$  & 59,726$^{\bigstar}$ & 3,812$^{\bigstar}$ & 13,227$^{\bigstar}$ & 19,281 & 1,037,115$^{\bigstar}$   & 456,658$^{\bigstar}$  & 1,582,875$^{\bigstar}$  & 4,748$^{\bigstar}$ & 5,890$^{\bigstar}$ & 6,708     \\
\hline    
SGDQ    & {\cellcolor[gray]{0.8}\textbf{239,953}} & {\cellcolor[gray]{0.8}\textbf{21,116}}  & {\cellcolor[gray]{0.8}\textbf{11,771}} & {\cellcolor[gray]{0.8}\textbf{54}} & {\cellcolor[gray]{0.8}\textbf{92}} & {\cellcolor[gray]{0.8}\textbf{114}}  & {\cellcolor[gray]{0.8}\textbf{ 833,402}} & {\cellcolor[gray]{0.8}\textbf{32,632}}  & {\cellcolor[gray]{0.8}\textbf{22,690}} & 117 & {\cellcolor[gray]{0.8}\textbf{172}} & {\cellcolor[gray]{0.8}\textbf{149}}     \\
\hline    
\hline    
\#Results   & 6,717,143 & 1,116  & 21 & 720 & 4 & 3   &   17,200,845   &   1,116    &  21   &   720   &   10 &   3        \\
\hline    
\hline    
\#\textit{OP\textsubscript{sp}}   & 579,364 & 76,943  & 1,543 & 27 & 22 & 7   &   591,271   &   77,233    &  1,454   &   27   &   21 &   7        \\
\hline    
\end{tabular}
\end{table*}

\begin{table*}[ht]
\captionsetup{belowskip=0pt,aboveskip=0pt}
\caption{Complex queries response time on SNIB-15000 dataset(in milliseconds).}
\centering
\label{tbl:complex-snib}
\scriptsize
\begin{tabular}{ | >{\centering\arraybackslash}m{0.50in}  ? >{\raggedleft\arraybackslash}m{0.6in} | >{\raggedleft\arraybackslash}m{0.6in} |>{\raggedleft\arraybackslash}m{0.6in} |>{\raggedleft\arraybackslash}m{0.6in} |>{\raggedleft\arraybackslash}m{0.6in} |>{\raggedleft\arraybackslash}m{0.6in} | >{\raggedleft\arraybackslash}m{0.6in}  | }
\hline
     &  \multicolumn{1}{c|}{S1} & \multicolumn{1}{c|}{S2}  & \multicolumn{1}{c|}{S3} & \multicolumn{1}{c|}{S4} & \multicolumn{1}{c|}{S5} & \multicolumn{1}{c|}{S6}  & \multicolumn{1}{c|}{S7}    \\
\hline    
RDF-3X  & 2,396,685 & 16,227  & 84,584 & 8,792 & N/A & 5,214,652 & 449,401     \\
\hline    
Virtuoso7    & 4,134,085 & 19,178  & 4,650 & 5,643 & 23,386 & 829,220 & 805,995      \\
\hline    
SGDQ    & {\cellcolor[gray]{0.8}\textbf{1,953,027}} & {\cellcolor[gray]{0.8}\textbf{12,454}}  & {\cellcolor[gray]{0.8}\textbf{1,021}} & {\cellcolor[gray]{0.8}\textbf{2,352}} & {\cellcolor[gray]{0.8}\textbf{3,922}} & {\cellcolor[gray]{0.8}\textbf{227,932}} & {\cellcolor[gray]{0.8}\textbf{66,407}}    \\
\hline    
\hline    
\#\textit{OP\textsubscript{sp}}     & 1,762,736 & 5,931  & 69 & 1,293 & 1,127 & 18,556 & 500      \\
\hline    
\end{tabular}
\end{table*}

\subsection{Complex queries on LUBM} \label{subsec:eval:lubm-complex}

\begin{figure}
\captionsetup{belowskip=0.5pt,aboveskip=0.5pt}
  \centering
  \includegraphics[height=0.08\textheight]{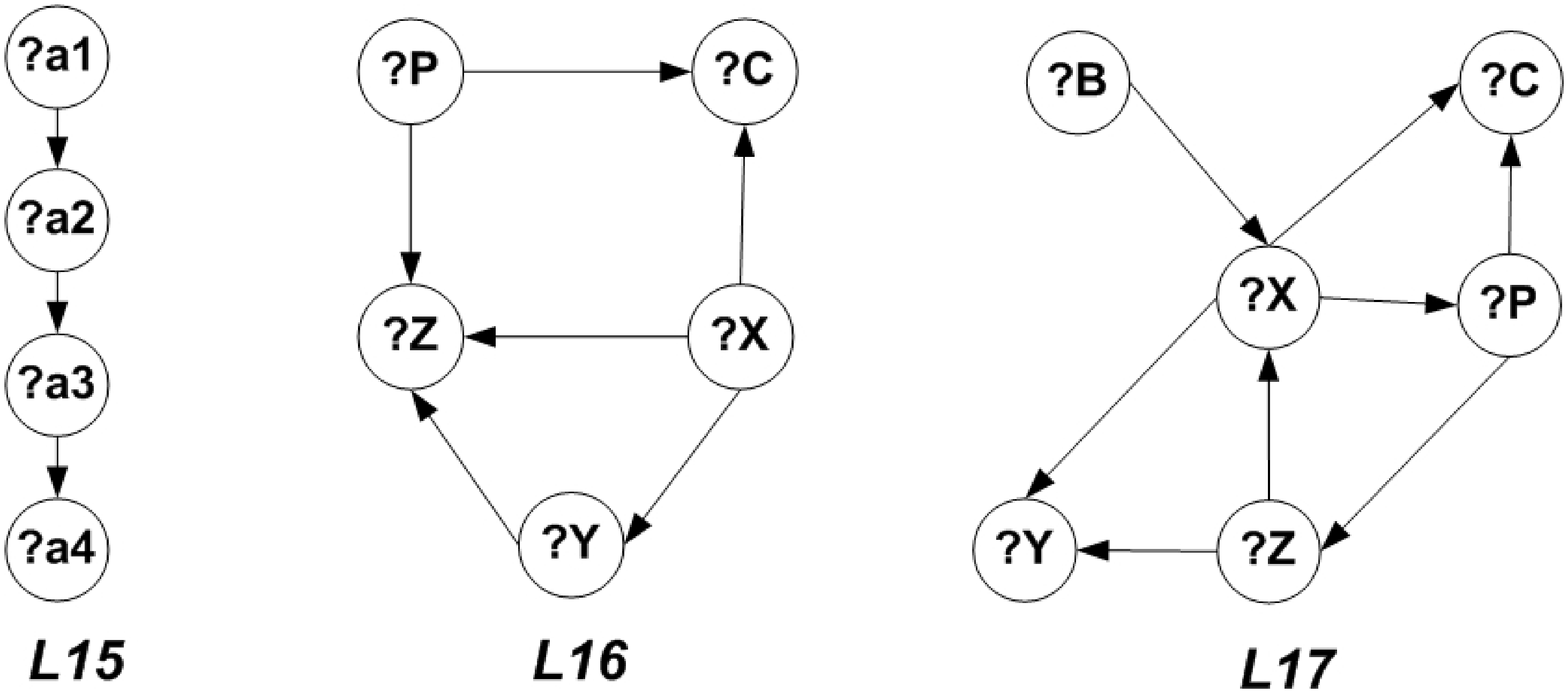}
\caption{The $G^Q_R$ of \textit{L15}, \textit{L16} and \textit{L17}}
\label{fig:complexlubmquery}
\end{figure}

As standard LUBM benchmark queries can not fully evaluate the performance of SGDQ, we further design 3 more complex queries on LUBM dataset. These queries show a more complex structural information. The $G^Q_R$ of \textit{L15-17} are shown as \textit{Figure} \ref{fig:complexlubmquery}. We make additional 3 queries by adding a highly selective \textit{A-Pattern} to each of them, and get \textit{L15}\textsubscript{sel}-\textit{L17}\textsubscript{sel} accordingly. These queries are listed in \textit{Appendix A}. Intuitively, the $\mathcal{TG}^Q$ of \textit{L15-17} have 3,4,4 vertices and 2,3,4 edges individually.

We report the response time of query on LUBM-8000 and LUBM-20480 dataset in \textit{Table} \ref{tbl:complex-lubm}, in both cold cache and warm cache mode. In fact, for complex query on large-scale dataset, this evaluation shows that only Virtuoso 7, which is a commercial-off-the-shelf system, is qualified. RDF-3X fails to return for \textit{L17} on LUBM-8000 and \textit{L16}, \textit{L17} on LUBM-20480. TripleBit returns with wrong number of results for all queries except \textit{L17}\textsubscript{sel}. For non-selective queries \textit{L15-17}, SGDQ consistently outperform others in both warm cache and cold cache mode.

To further investigate this problem, we record the total counts of \textit{OP\textsubscript{sp}} operations for each query processing. We can see that \textit{L15} generates more \textit{OP\textsubscript{sp}}, as each sub-queries of \textit{L15} is a simplex pattern pattern. SGDQ implements such \textit{OP\textsubscript{sp}} and \textit{OP\textsubscript{ipj}} at the magnitude of approximately 1 ms, this contributes to the overall query performance promotion. Compare the result of \textit{L15} on LUBM-8000 and LUBM-20480, we can see that as the data size increase, the average cost of \textit{OP\textsubscript{sp}} increase accordingly. This is due to expansion of partition size. Although \textit{L17} generates far more less \textit{OP\textsubscript{sp}}, but most of its \textit{OP\textsubscript{sp}} are triangles, and the average execution time is at a magnitude of 10 to 20 ms. This evaluation shows the efficiency of \textit{OP\textsubscript{sp}} in LUBM dataset.

\subsection{Complex queries on SNIB} \label{subsec:eval:snib-complex}

We also evaluated the query performance on a SNIB benchmark dataset, which is generated as a social network of 15000 users. SNIB dataset presents a more connected graph structure compared with LUBM dataset, thus we can introduce more complex queries to evaluate the effect of SGDQ. We define 7 SNIB queries as listed in \textit{Appendix B}. Their $G^Q_R$ are shown as \textit{Figure} \ref{fig:complexlubmquery}.  We report the response time of query on SNIB-15000 dataset in \textit{Table} \ref{tbl:complex-snib}. All queries were executed in warm cache mode. As TripleBit is inclined to to return wrong number of results, we do not select it in this evaluation. In essence, SGDQ consistently outperforms all competitors as the queries are more complex.

\begin{figure}
  \centering
  \includegraphics[height=0.15\textheight]{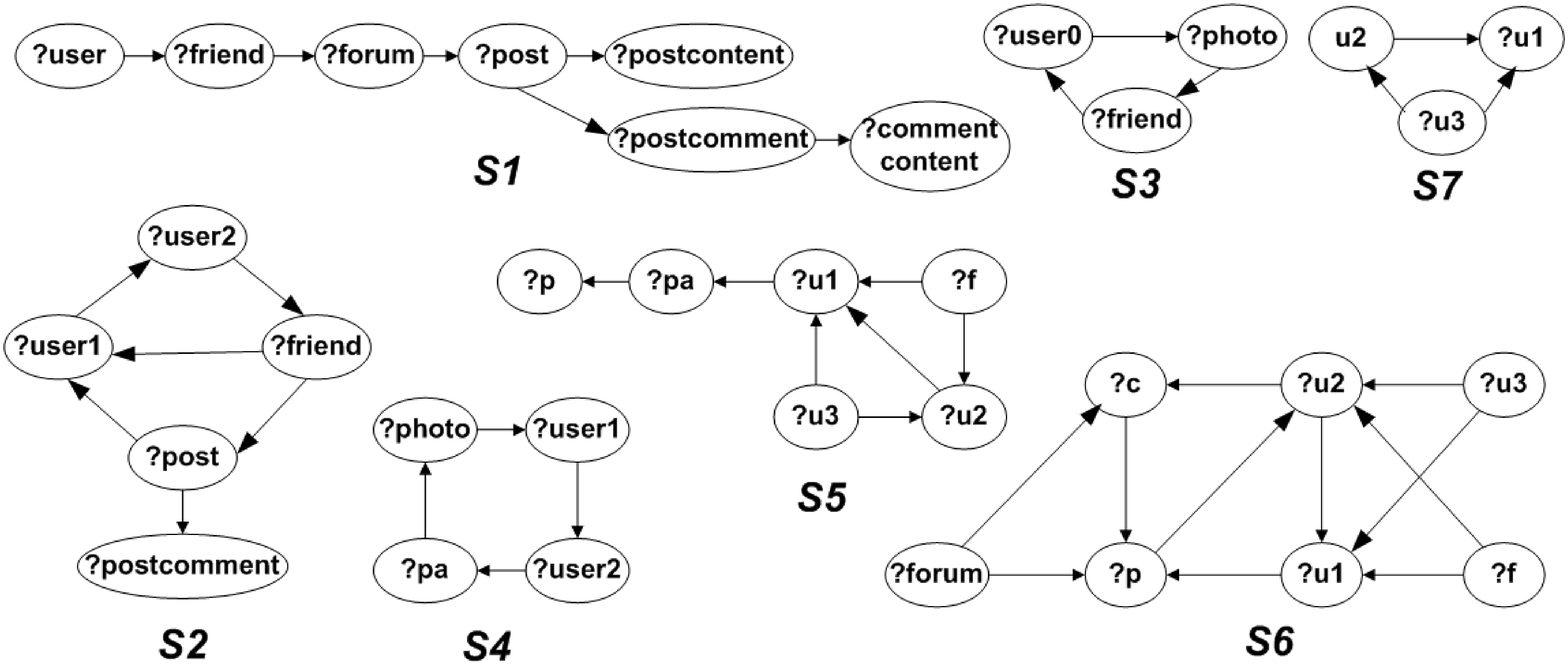}
\caption{The $G^Q_R$ of 7 SNIB queries}
\label{fig:complexsnibquery}
\end{figure}

To elaborate, query \textit{S1} and \textit{S4} shows a similar characteristic. All the  \textit{OP\textsubscript{sp}} are \textit{Type-I} sub-query. This cause the number of  \textit{OP\textsubscript{sp}} grows exponentially with the number of edges in $G^Q_R$. SGDQ shows a superior performance due to the efficiency of \textit{OP\textsubscript{sp}}. Query \textit{S3} and \textit{S7} have a triangle $G^Q_R$, and their $\mathcal{TG}^Q$ contains only one vertex. They differ in that \textit{S7} does not have a selective \textit{A-Pattern} as \textit{S3}. \textit{S7} needs to explore all partitions. Like in LUBM evaluation, SGDQ also performs well on such queries. Query \textit{S2}, \textit{S5} and \textit{S6} feature the processing of complex \textit{OP\textsubscript{ipj}} between \textit{Type-II} sub-queries. RDF-3X performs the worst due to excessive number of joins, even fails for \textit{S5}. The superior performance of SGDQ on these queries shows that all \textit{OP\textsubscript{ipj}} are also efficiently processed in SGDQ. These complex queries benefit from the simplification of query processing using query decomposition and data partition.

%
%
%
%
\section{Conclusion}\label{sec-conclusion}

We have presented SGDQ, a \textit{\textbf{S}ummary \textbf{G}raph \textbf{D}riven \textbf{Q}uery} framework for RDF query processing. SGDQ is based on efficient execution of sub-queries on partitioned data. It utilizes the effectiveness of subgraph matching in representing the inter-partition query processing, and avoids the problem of maintaining huge intermediate results in canonical relational query methods. To enable SGDQ, we introduced specialized physical data layout and implemented efficient operations in SGDQ. We have performed an extensive evaluation on the query performance of SGDQ compared with three competitive RDF systems. We also noticed that in current work, the exhaustive exploration of neighbourhood in summary graph match introduces many useless inter-partition joins. In the future, we plan to investigate the mechanism to prune unnecessary partitions by maintaining fine-grained inter-partition data in the summary graph.

%
%
%
%


%
%
%
%
\bibliographystyle{IEEEtran}
\bibliography{IEEEabrv,icde2016-ref}

%
%
%
%
\section*{Appendix}\label{sec-appendix}
\begin{scriptsize}
\subsection{LUBM Queries} \label{appendix-lubmquery}

\vspace{-1em} \noindent PREFIX rdf: $\langle$http://www.w3.org/1999/02/22-rdf-syntax-ns\#$\rangle$

\vspace{-1em} \noindent PREFIX rdfs: $\langle$http://www.w3.org/2000/01/rdf-schema\#$\rangle$

\vspace{-1em} \noindent PREFIX ub: $\langle$http://swat.cse.lehigh.edu/onto/univ-bench.owl\#$\rangle$

\vspace{-1em} \noindent \textbf{L1-L14:}Same as LUBM Q1-Q14 used in \cite{subgraph-vldb2015} respectively.

\vspace{-1em} \noindent \textbf{L15:}SELECT ?a1 ?a2 ?a3 ?a4 WHERE\{?a1 ub:advisor ?a2. ?a2 ub:worksFor ?a3. ?a3 ub:subOrganizationOf ?a4. ?a4 rdf:type ub:University. ?a1 rdf:type ub:GraduateStudent. ?a2 rdf:type ub:FullProfessor. ?a3 rdf:type ub:Department. \}

\vspace{-1em} \noindent \textbf{L15\textsubscript{sel}:}SELECT ?a1 ?a2 ?a3 ?a4 WHERE\{?a1 ub:advisor ?a2. ?a2 ub:worksFor ?a3. ?a3 ub:subOrganizationOf ?a4. ?a1 rdf:type ub:GraduateStudent. ?a2 rdf:type ub:FullProfessor. ?a3 rdf:type ub:Department. ?a4 ub:name "University7". \}

\vspace{-1em} \noindent \textbf{L16:}SELECT ?X ?Y ?Z WHERE\{?X rdf:type ub:GraduateStudent. ?Y rdf:type ub:University.
 ?Z rdf:type ub:Department.
 ?C rdf:type ub:GraduateCourse.
 ?X ub:takesCourse ?C.
 ?P rdf:type ub:FullProfessor.
 ?P ub:teacherOf ?C.
 ?P ub:worksFor ?Z.
 ?X ub:memberOf ?Z.
 ?Z ub:subOrganizationOf ?Y.
 ?X ub:undergraduateDegreeFrom ?Y\}

\vspace{-1em} \noindent \textbf{L16\textsubscript{sel}:}SELECT ?X ?Y ?Z WHERE\{?X rdf:type ub:GraduateStudent.
 ?Z rdf:type ub:Department.
 ?C rdf:type ub:GraduateCourse.
 ?X ub:takesCourse ?C.
 ?P rdf:type ub:FullProfessor.
 ?P ub:teacherOf ?C.
 ?P ub:worksFor ?Z.
 ?X ub:memberOf ?Z.
 ?Z ub:subOrganizationOf ?Y.
 ?X ub:undergraduateDegreeFrom ?Y.
 ?Y ub:name "University6". \}

\vspace{-1em} \noindent \textbf{L17:}SELECT ?X ?Y ?P ?B ?C WHERE\{?X rdf:type ub:GraduateStudent.
 ?X rdf:type ub:TeachingAssistant.
 ?X ub:advisor ?P.
 ?X ub:takesCourse ?C.
 ?X ub:memberOf ?Z.
 ?X ub:undergraduateDegreeFrom ?Y.
 ?Y rdf:type ub:University.
 ?C rdf:type ub:GraduateCourse.
 ?Z rdf:type ub:Department.
 ?Z ub:subOrganizationOf ?Y.
 ?P rdf:type ub:AssociateProfessor.
 ?P ub:worksFor ?Z.
 ?P ub:teacherOf ?C.
 ?B rdf:type ub:Publication.
 ?B ub:publicationAuthor ?X. \}

\vspace{-1em} \noindent \textbf{L17\textsubscript{sel}:}SELECT ?X ?Y ?P ?B ?C WHERE\{?X rdf:type ub:GraduateStudent.
 ?X rdf:type ub:TeachingAssistant.
 ?X ub:advisor ?P.
 ?X ub:takesCourse ?C.
 ?X ub:memberOf ?Z.
 ?X ub:undergraduateDegreeFrom ?Y.
 ?Y rdf:type ub:University.
 ?Y ub:name "University786" .
 ?C rdf:type ub:GraduateCourse.
 ?Z rdf:type ub:Department.
 ?Z ub:subOrganizationOf ?Y.
 ?P rdf:type ub:AssociateProfessor.
 ?P ub:worksFor ?Z.
 ?P ub:teacherOf ?C.
 ?B rdf:type ub:Publication.
 ?B ub:publicationAuthor ?X. \}


\subsection{SNIB Queries} \label{appendix-snibquery}

\vspace{-1em} \noindent PREFIX foaf: $\langle$http://xmlns.com/foaf/0.1/$\rangle$

\vspace{-1em} \noindent PREFIX dc: $\langle$http://purl.org/dc/elements/1.1/$\rangle$

\vspace{-1em} \noindent PREFIX sioc: $\langle$http://rdfs.org/sioc/ns\#$\rangle$

\vspace{-1em} \noindent PREFIX sioct: $\langle$http://rdfs.org/sioc/type\#$\rangle$

\vspace{-1em} \noindent PREFIX sib: $\langle$http://www.ins.cwi.nl/sib/vocabulary/$\rangle$

\vspace{-1em} \noindent PREFIX rdf: $\langle$http://www.w3.org/1999/02/22-rdf-syntax-ns\#$\rangle$

\vspace{-1em} \noindent PREFIX dbp: $\langle$http://dbpedia.org/resource/$\rangle$

\vspace{-1em} \noindent \textbf{S1:}SELECT ?user ?commentcontent ?commentdate WHERE\{?user foaf:knows ?friend.
 ?user rdf:type sib:User.
 ?friend rdf:type sib:User.
 ?friend sioc:moderator\_of ?forum.
 ?post rdf:type sioc:Post.
 ?forum sioc:container\_of ?post.
 ?post sioc:content ?postcontent.
 ?post sib:hashtag dbp:Island .
 ?post sioc:container\_of ?postcomment.
 ?postcomment rdf:type sioc:Item.
 ?postcomment sioc:content ?commentcontent.
 ?postcomment dc:created ?commentdate. \}

\vspace{-1em} \noindent \textbf{S2:}SELECT ?user1 ?user2 ?friend WHERE\{?user1 foaf:knows ?user2.
 ?user2 foaf:knows ?friend .
 ?friend foaf:knows ?user1 .
 ?user1 rdf:type sib:User .
 ?user2 rdf:type sib:User .
 ?friend rdf:type sib:User .
 ?friend sioc:creator\_of ?post.
 ?post rdf:type sioc:Post .
 ?post sioc:content ?postcontent.
 ?post sib:hashtag dbp:Arrow .
 ?post sib:liked\_by ?user1\}
		
\vspace{-1em} \noindent \textbf{S3:}SELECT ?user0 ?friend ?photo WHERE\{?user0 rdfs:type sib:User.
 ?friend rdf:type sib:User.
 ?photo rdf:type sib:Photo .
 ?friend foaf:knows ?user0.
 ?photo sib:usertag ?friend.
 ?photo dbp:location "Germany".
 ?user0 sib:liked\_by ?photo.\}

\vspace{-1em} \noindent \textbf{S4:}SELECT ?photo WHERE\{?photo rdfs:type sib:Photo.
 ?user1 rdf:type sib:User .
 ?user2 rdf:type sib:User .
 ?pa rdfs:type sioct:ImageGallery.
 ?photo sib:usertag ?user1.
 ?photo dbp:location "Crimea".
 ?user1 foaf:knows ?user2.
 ?pa sioc:container\_of ?photo.
 ?user2 sioc:creator\_of ?pa. \}
		
\vspace{-1em} \noindent \textbf{S5:}SELECT ?u1 ?u2 ?p WHERE\{?f sib:memb ?u1.
 ?f sib:memb ?u2.
 ?u2 foaf:knows ?u1.
 ?u3 foaf:knows ?u2.
 ?u3 foaf:knows ?u1.
 ?u1 sioc:creator\_of ?pa.
 ?u1 rdf:type sib:User.
 ?u2 rdf:type sib:User.
 ?u3 rdf:type sib:User.
 ?pa rdf:type sioct:ImageGallery.
 ?pa sioc:container\_of ?p.
 ?p rdf:type sib:Photo.
 ?p dbp:location "Germany".\}

\vspace{-1em} \noindent \textbf{S6:}SELECT ?u1 ?u2 WHERE\{?f sib:memb ?u1.
 ?f sib:memb ?u2.
 ?u2 foaf:knows ?u1.
 ?u3 foaf:knows ?u2.
 ?u3 foaf:knows ?u1.
 ?u1 rdf:type sib:User.
 ?u2 rdf:type sib:User.
 ?u3 rdf:type sib:User.
 ?u1 sioc:creator\_of ?p.
 ?p rdf:type sioc:Post.
 ?forum sioc:container\_of ?p.
 ?forum sioc:container\_of ?c.
 ?p sib:hashtag dbp:Island.
 ?c rdf:type sioc:Item.
 ?u2 sioc:creator\_of ?c.
 ?c sioc:reply\_of ?p.
 ?p sib:liked\_by ?u2. \}

\vspace{-1em} \noindent \textbf{S7:}SELECT ?u1 ?u2 ?p WHERE\{?u2 foaf:knows ?u1.
 ?u3 foaf:knows ?u2.
 ?u3 foaf:knows ?u1. \}

\end{scriptsize}

\end{document}